\documentclass[10pt]{aastex62}

\usepackage{xcolor}
\usepackage[utf8]{inputenc}
\usepackage[english]{babel}
\usepackage{hyperref}

\newenvironment{tight_enumerate}{
\begin{enumerate}
  \setlength{\itemsep}{0pt}
  \setlength{\parskip}{0pt}
}{\end{enumerate}}

\shorttitle{Demographics in SDSS}
\shortauthors{SDSS-IV COINS}

\begin{document}
\title{SDSS-IV from 2014 to 2016: A Detailed Demographic Comparison over Three Years}

\author[0000-0002-2262-8240]{Amy M. Jones}
\affiliation{Space Telescope Science Institute, Baltimore, MD, 21218, USA}
\affiliation{Physics and Astronomy, University of Alabama, Tuscaloosa, AL, United States}
\author[0000-0002-1691-8217]{Rachael L. Beaton}
\affiliation{Space Telescope Science Institute, Baltimore, MD, 21218, USA}
\affiliation{Department of Astrophysical Sciences, 4 Ivy Lane, Princeton University, Princeton, NJ 08544}
\affiliation{The Observatories of the Carnegie Institution for Science, 813 Santa Barbara St., Pasadena, CA~91101}
\author[0000-0002-4289-7923]{Brian A. Cherinka}
\affiliation{Space Telescope Science Institute, Baltimore, MD, 21218, USA}
\author[0000-0003-0846-9578]{Karen L. Masters}
\affiliation{Department of Physics and Astronomy, Haverford College, 370 Lancaster Avenue, Haverford, PA 19041, USA}
\author[0000-0001-8808-0073]{Sara Lucatello}
\altaffiliation{Anna Boyksen Fellow}
\affiliation{INAF Osservatorio Astronomico di Padova, vicolo dell'Osservatorio 5, 35122 Padova, Italy}
\affiliation{Institute for Advanced Studies, TUM, Lichtenbergstrasse 2 a 85748 Garching, Germany}
\author{Aleksandar M. Diamond-Stanic}
\affiliation{Department of Physics and Astronomy, Bates College, Lewiston, ME 04240, USA}
\author[0000-0002-5469-5149]{Sarah A. Bird}
\affiliation{College of Science, China Three Gorges University, Yichang 443002, People’s Republic of China}
\affiliation{Center for Astronomy and Space Sciences, China Three Gorges University, Yichang 443002, People’s Republic of China}
\affiliation{CAS Key Laboratory of Optical Astronomy, National Astronomical Observatories, Chinese Academy of Sciences, Beijing 100101, People’s Republic of China}
\affiliation{Shanghai Astronomical Observatory, Chinese Academy of Sciences, 80 Nandan Road, Shanghai 200030, People’s Republic of China}
\author{Michael R.~Blanton}
\affiliation{Center for Cosmology and Particle Physics,
Department of Physics,
New York University,
726 Broadway Rm. 1005,
New York, NY 10003, USA}
\author[0000-0001-6476-0576]{Katia Cunha}
\affiliation{Steward Observatory, University of Arizona, Tucson, AZ 85721, USA}
\affiliation{Institut d'Astrophysique de Paris, CNRS and Sorbonne Université, 75014 Paris, France}
\affiliation{Observat\'orio Nacional MCTI, S\~ao Crist\'ov\~ao, Rio de Janeiro, Brazil}
\author[0000-0002-5454-8157]{Emily E. Farr}
\affiliation{Laboratory for Atmospheric and Space Physics, University of Colorado, Boulder CO 80309, USA}
\author[0000-0002-3101-5921]{Diane Feuillet}
\affiliation{Lund Observatory, Department of Geology,
S\"olvegatan 12, SE-223\,62 Lund,  Sweden}
\author[0000-0002-0740-8346]{Peter M. Frinchaboy}
\affiliation{Department of Physics and Astronomy, Texas Christian University, TCU Box 298840 Fort Worth, TX 76129, USA}
\author{Alex Hagen}
\affiliation{Qualcomm, Inc., 5775 Morehouse Drive, San Diego, CA 92121, USA}
\author{Karen Kinemuchi}
\altaffiliation{Current affiliation: Frontier Technology, Inc, 100 Cummings Center, Suite 450G, Beverly MA 01915}
\affiliation{Apache Point Observatory/NMSU, PO Box 59, Sunspot, NM 88349, USA}
\author[0000-0002-6463-2483]{Britt Lundgren}
\affiliation{Department of Physics and Astronomy, University of North Carolina Asheville, Asheville, North Carolina, 28804, USA}
\author[0000-0002-4775-7292]{Mariarosa L. Marinelli}
\affiliation{Space Telescope Science Institute, Baltimore, MD, 21218, USA}
\author{Adam D.\ Myers}
\affil{Department of Physics and Astronomy, University of Wyoming, Laramie, WY, 82071, USA}
\author[0000-0002-1379-4204]{Alexandre Roman-Lopes}
\affiliation{Department of Astronomy - Universidad de La Serena - Av. Juan Cisternas, 1200 North, La Serena, Chile}
\author{Ashley J. Ross}
\affiliation{Center for Cosmology and AstroParticle Physics, The Ohio State University, 191 West Woodruff Avenue, Columbus, OH 43210, USA}
\author[0000-0003-2486-3858]{Jos\'e~R.~S\'anchez-Gallego}
\affiliation{Department of Astronomy, University of Washington, Box 351580, Seattle, WA 98195, USA}
\author[0000-0002-7224-7702]{Sarah J. Schmidt}
\affiliation{Leibniz Institute for Astrophysics Potsdam (AIP), An der Sternwarte 16, 14482 Potsdam, Germany}
\author[0000-0002-4989-0353]{Jennifer Sobeck}
\affiliation{Canada-France-Hawaii Telescope, Waimea, HI, 96743, USA}
\author[0000-0002-3481-9052]{Keivan G.\ Stassun}
\affiliation{Department of Physics and Astronomy, Vanderbilt University, Nashville, TN 37235, USA}
\author[0000-0002-4818-7885]{Jamie Tayar}
\affiliation{Department of Astronomy, University of Florida,  USA }
\author[0000-0003-3841-1836]{Mariana Vargas-Magaña}
\affiliation{Instituto de Física, Universidad Nacional Autónoma de México, Apdo. Postal 20-364, 01000, D.F, México}
\author[0000-0001-7828-7257]{J.~C.~Wilson}
\affiliation{Astronomy Department, University of Virginia,
Charlottesville, VA 22901, USA}
\author[0000-0001-6761-9359]{Gail Zasowski}
\affiliation{Department of Physics \& Astronomy, University of Utah, Salt Lake City, UT 84112, USA}

\begin{abstract}

The Sloan Digital Sky Survey (SDSS) is one of the largest international astronomy organizations. We present demographic data based on surveys of its members from 2014, 2015 and 2016, during the fourth phase of SDSS (SDSS-IV). We find about half of SDSS-IV collaboration members were based in North America, a quarter in Europe, and the remainder in Asia and Central and South America. Overall, $26-36$\% are women (from 2014 to 2016), up to 2\% report non-binary genders. $11-14$\% report that they are racial or ethnic minorities where they live. The fraction of women drops with seniority, and is also lower among collaboration leadership. Men in SDSS-IV were more likely to report being in a leadership role, and for the role to be funded and formally recognized. SDSS-IV collaboration members are twice as likely to have a parent with a college degree, than the general population, and are ten times more likely to have a parent with a PhD. This trend is slightly enhanced for female collaboration members. Despite this, the fraction of first generation college students (FGCS) is significant (31\%). This fraction increased among collaboration members who are racial or ethnic minorities ($40-50$\%), and decreased among women ($15-25$\%). SDSS-IV implemented many inclusive policies and established a dedicated committee, the Committee on INclusiveness in SDSS (COINS). More than 60\% of the collaboration agree that the collaboration is inclusive; however, collaboration leadership more strongly agree with this than the general membership. In this paper, we explain these results in full, including the history of inclusive efforts in SDSS-IV. We conclude with a list of suggested recommendations based on our findings, which can be used to improve equity and inclusion in large astronomical collaborations, which we argue is not only moral, but will also optimize their scientific output.

\end{abstract}

\keywords{Surveys (1671), Sociology of astronomy (1470)}

\section{Introduction}

Large multi-institutional and multi-national collaborations are now a major component of astronomical research.
These organizational structures allow the combination of resources and the sharing of costs associated with ever more complex and ambitious experiments.  They also facilitate the exchange of expertise and knowledge among the member institutions and participants.
This change in the landscape of observational astronomy leads to new and unique challenges regarding the integration and blending of different cultures and backgrounds, as well as the conscious creation of inclusive environments needed for all scientists and staff to thrive.
Cultivating project culture can be especially challenging when there is limited face-to-face interaction between participants and no physical environment that they share. In addition, the Covid-19 pandemic and increase in hybrid work further restricted in person collaborations.

The Sloan Digital Sky Survey (SDSS) continues to be one of the largest international astronomy collaborations. 
For over two decades, SDSS has brought together scientific and technical personnel at all career stages, including undergraduate students, graduate students, postdoctoral researchers, junior and senior staff with temporary and permanent positions.
We focus on SDSS during its fourth generation \citep[SDSS-IV]{blanton_2017}, which ran from 2014 to 2020 and involved members from over 60 institutions, located in 18 countries, spanning five continents, and around 1500 active accounts on the internal wiki.  SDSS-IV had three main surveys, namely the Apache Point Observatory Galactic Evolution Experiment 2 (APOGEE-2), Mapping Nearby Galaxies at Apache Point Observatory (MaNGA), and the extended Baryon Oscillation Spectroscopic Survey (eBOSS).  APOGEE-2 obtained near-infrared spectra of hundreds of thousands of stars in the Milky Way.  MaNGA collected optical integral field spectroscopy of ten thousand nearby galaxies.  eBOSS mapped the galaxy, quasar, and neutral gas distributions at intermediate redshifts to constrain cosmology.  eBOSS also had two subprograms, the SPectroscopic IDentification of eROSITA Sources (SPIDERS), investigating X-ray Active Galactic Nuclei and galaxies in X-ray clusters, and the Time Domain Spectroscopic Survey (TDSS), obtaining spectra of variable sources.  Additionally, the MaNGA stellar library (MaStar) provided an optical stellar library covering a wide range of stellar parameters.  These surveys used the 2.5\,m Sloan Foundation Telescope at Apache Point Observatory, U.S.A.  In 2017, APOGEE-2 started observing with a second near-infrared spectrograph on the 2.5\,m du Pont Telescope at Las Campanas Observatory, Chile.  

SDSS-IV had regularly scheduled public data releases, with the first in SDSS-IV being Data Release 13 in July 2016 \citep{DR13}. Funding for SDSS has always involved a component from the Sloan Foundation\footnote{The Sloan Foundation is a not-for-profit grant-making institution which supports scientific research; \url{https://sloan.org }} (usually about 25\%), with the bulk coming from institutional buy-in (about 75\%) and a small fraction from federal grants. The annual SDSS collaboration meeting brings together roughly 10\% of the membership to a location that rotates among geographical regions.  SDSS is currently in its fifth generation \citep[SDSS-V]{Kollmeier2017}, which started in 2020. 
Over its history, SDSS has profoundly influenced modern astronomy, not only in paving the way for large astronomical collaborations and demonstrating the power of open science, but also in its management styles and policies that can be considered a factor in its broad success \citep[see discussions in][]{finkbeiner2010grand,Blanton_2019,Gunn_2020,michelson2020philanthropy,MasterSkyTel}. 

For a collaboration of the size and influence of SDSS, it is essential to establish and maintain a climate of inclusiveness, where every member is able to contribute fully and offered fair and equitable opportunities for career advancement \citep[e.g.][and references therein]{hunter_2007}.
There is ample evidence that diverse working groups, especially in science, technology, engineering, and mathematics (STEM), not only foster innovation and creativity, but produce more engaged and productive team-environments. For an overview that is contextually focused on astronomical observatories, see \citet[][and references therein]{shugart_2018}. 

SDSS provides annual reports to the Sloan Foundation. In the early 2010s, the Sloan Foundation recommended that SDSS-III evaluate the demographics of the collaboration; of particular interest at that time was the participation of women in leadership positions \citep[see discussion in][]{Blanton_2019,michelson2020philanthropy}.
This resulted in the implementation of a demographics survey that was first administered in 2014 to SDSS-III and SDSS-IV members and is presented in \citet{lundgren_2015}. 
This effort was the first of its kind at the scale an astronomical collaboration\footnote{This was communicated in a confidential report provided to SDSS.}. 
SDSS-IV continued conducting regular demographic surveys. \autoref{app:history} provides a more detailed account of the history of inclusion efforts in SDSS-IV, and \autoref{app:current} discusses policies enacted in SDSS-IV based on an American Institute of Physics (AIP) report that evaluated the climate of SDSS.

Larger, national and international astronomical organizations also perform regular demographic surveys:
the American Astronomical Society (AAS) in 2013 \citep{anderson_2013}, and 2016 \citep{pold_2017}
\footnote{The 2016 study is available: \url{https://aas.org/comms/demographics-committee}}(specialized surveys are conducted more often), the Royal Astronomical Society (RAS) in 2015 \citep{Massey_2015} and 2017 \citep{Massey_2017}, and the International Astronomical Union (IAU) has some annual demographics reports available\footnote{We will use data from 2016 that is found in attachments to this URL \url{https://www.iau.org/news/announcements/detail/ann16020/}} as well as a monthly update on the current membership status. 
Through its Working Group on Inclusion and Gender \footnote{\url{https://sochias.cl/actividades/grupo-de-trabajo-en-inclusion-y-genero/}}, the Sociedad Chilena de Astronomia undertook its first survey in 2022.
Physical societies perform demographic surveys that include astronomers, for example the results from a recent survey by the AIP is given in \citet{porter_2019}.

In addition to surveys done by organizations, there are other sources of demographic data in astronomy. Some astronomical observatories maintain demographic data, but these focus on individuals that were hired into positions and the analysis is typically in regards to its hiring practices \citep[][and references therein]{shugart_2018}.
Reports on the demographics of proposals presented to Telescope Allocation Committees (TACs) or grant committees, and those awarded time are also becoming more common \citep{Reid2014_HSTGender,Patat_2016,Spekkens2018,Piccialli2020,Carpenter_2020}. In some cases, these have led to policy modifications resulting in substantive change, such as the Hubble Space Telescope TAC process \citep[see e.g.,][]{Johnson_2020,2021BAAS...53d.010A}.
SDSS differs from national-level astronomical societies and astronomical observatories in that the average collaboration member is not hired by SDSS and can opt-in or buy-in to participate.\footnote{We note that there are a number of ``formal'' positions at the graduate, postdoctoral, and staff/administration levels in SDSS. The bulk of these individuals are predominantly hired or admitted, by their respective institutions instead of by SDSS-IV, itself.} 

The importance of collecting self-reported demographic data for both internal and external evaluation cannot be overstated. 
One aspect of its importance has to do with biased perceptions of demographic factors. For example, the prevalence of female Chief Executive Officers (CEOs) of major companies is perceived by executives, regardless of gender, to be between 20\% to 25\%, but is nearly 4-fold smaller at 8\% \citep{ceo_2015}. 
Perceptions of ``fair'' gender representation in the print media, in film/television, or even in spoken conversations are more commonly identified by men, when the actual male:female ratio is as high as 4:1 \citep[][among others]{cutler_1990,len-rios_2005,smith_2015}.
Often our interpretations are influenced by``implicit bias'' wherein societal stereotypes and other unconscious biases impact our perceptions \citep[for a summary with respect to astronomy, see][]{knezek_2017}.
Thus, relying on anecdotal or perceptional evidence may lead to biased inferences and ineffectual solutions; this is particularly true for managerial decision-making.
Indeed, formal training in how to collect and interpret demographic data has been recommended as a requirement for effective leadership in astronomy \citep{brinkworth_2016}.

Likewise, perceptions of demographic diversity also requires self-reported data. 
Perceptions of diversity are more likely to focus on surface-level diversity conveyed via visible differences, rather than deep-diversity, a term that defines differences in  world-views, experiences, and thinking systems that are tied to increased creativity and innovation \citep{https://doi.org/10.1002/job.2362}.
While surface- and deep-diversity classifications do correlate strongly in some situations, in others --- particularly in global, multinational teams --- the distinction may be less clear.
Moreover, in global virtual teams, like SDSS-IV, the means of promoting cultural cohesion depend on the recognition and understanding of cultural differences in managing teams that do not interact regularly in person \citep{Stahl2010UnravelingTE}. 
Businesses that have diverse management have been shown to outperform industry averages by 36\%, with the focus on management diversity being interpreted as not just actions that the companies take but also as being impactful for the methods through which teams work within the company \citep{diversitywins}. For SDSS to work towards having diverse management, it requires self-reported demographics data of its leadership and members. 

This paper presents demographic data from SDSS that was obtained in surveys conducted in 2015 and 2016; some results from the 2015 survey were discussed in \citet{lucatello_2017}. 
These previously unpublished data are compared directly with those from the 2014 survey \citep[e.g.,][]{lundgren_2015} with the explicit goal of evaluating {\it if} and {\it how} the demographics have changed over time. 

The demographic survey itself is described in \autoref{sec:description}. 
\autoref{sec:overalldemo} presents the overall demographic portrait of SDSS for the three survey years and compares to data from other astronomical organizations, when available.
Specifically, \autoref{sec:gender} describes the gender balance with academic age and \autoref{sec:edubkgd} with socio-economic status. 
\autoref{sec:leadership} reports on the gender balance of SDSS leadership and explores the intersection of these statistics with policies around filling leadership positions.
\autoref{sec:surveyclimate} explores respondents' perceptions of the collaboration climate. 
Findings and recommendations from the paper are summarized in \autoref{sec:summary}. 
Additional contextual information is provided in a set of appendices, specifically: 
    \autoref{app:history} provides a history of inclusion efforts in SDSS, 
    \autoref{app:current} describes specific polices enacted in SDSS-IV, and 
    \autoref{app:survey} provides the complete 2016 Demographic Survey.

\section{Description of the Demographic Survey} \label{sec:description}

The demographic survey included questions in the following categories:
\begin{tight_enumerate}
    \item Career Information,
    \item Experience within SDSS,
    \item Demographic Information, and
    \item Leadership Status in SDSS.
\end{tight_enumerate}
To maintain consistency with the 2014 SDSS-IV Demographics Survey \citep{lundgren_2015}, questions from 2014 regarding \texttt{Career Information} and \texttt{Demographic Information} were unchanged in the 2015 and 2016 surveys.  
Feedback from members suggested the inclusion of additional questions in 2015 and 2016 within \texttt{Demographic Information} and new information was collected with regards to: (i) identification with the LGBT community, (ii) disability status, (iii) partnership status, (iv) family status, and (v) parental educational achievement.
Additionally, the new section \texttt{Experience with SDSS} was added to trace specific information about the climate of SDSS-IV.
The full 2016 survey is provided in \autoref{app:survey}.

All questions in the survey were voluntary and optional, and respondents were reminded of this on each page of the survey. 
Beginning in the 2016 survey, \texttt{prefer not to answer} was added as an option on each question to further clarify that all questions are optional.
We did not ask nor collect any identifying information from the respondents in 2015 and 2016 (e.g., email or IP addresses);
in 2014 there was an optional free response to collect an email address for one-on-one follow-up. 
For all three survey years, the time of submission was recorded, which helps us understand response rates through the time period the survey was accepting responses.
The data collected was only made available to committee members who administered the survey. Anyone joining the committee later could only see the aggregate data. 
Within the context of this publication, if any given combination of demographic information could make any individual identifiable, it has been excluded from any visualization and the numbers are not reported.

The link to the survey was sent via email to the SDSS-IV general email list to reach all SDSS members with an account. Even with the new questions in the 2015 and 2016 surveys, the surveys could still be completed in less than 5 minutes.
To advertise each year's surveys, we enlisted the SDSS-IV Director and Spokesperson to circulate emails to encourage participation.  

The demographics surveys were completed by 240, 351, and 246  SDSS-IV members in 2014, 2015, and 2016, respectively.
As with the 2014 survey, we use the number of collaboration wiki subscribers to define an upper bound on the total number of SDSS-IV members, which was 1485 at the time our detailed analysis began in 2018\footnote{It is possible to use the SDSS Database to determine the number of wiki accounts closer to the time that the surveys were conducted, however it is complicated and this is only used as an upper limit.}. 
Wiki accounts are provided to SDSS collaboration members and to external collaborators. Accounts remain open even if members change institutions. Thus, the number of wiki accounts serves as an upper bound on the number of members active in SDSS at any given time. 
Using $N_\mathrm{wiki}$=1485 as an upper bound on the number of SDSS members suggests minimum response rates to the demographic surveys of 17, 24, and 17\% in 2014, 2015, and 2016, respectively. 
We note in 2014 using a very contemporaneous wiki accounting, found a response rate of 46\% compared to our lower limit of 17\% \citep{lundgren_2015}. The higher number of responses in 2015 may be due to a campaign to reach a high-response rate.  This included opening the survey soon after the publication of the 2014 survey \citep{lundgren_2015}, which may have motivated collaboration members to fill out a survey they could see would be used. 

\begin{figure}
    \centering
    \includegraphics[width=0.73\textwidth]{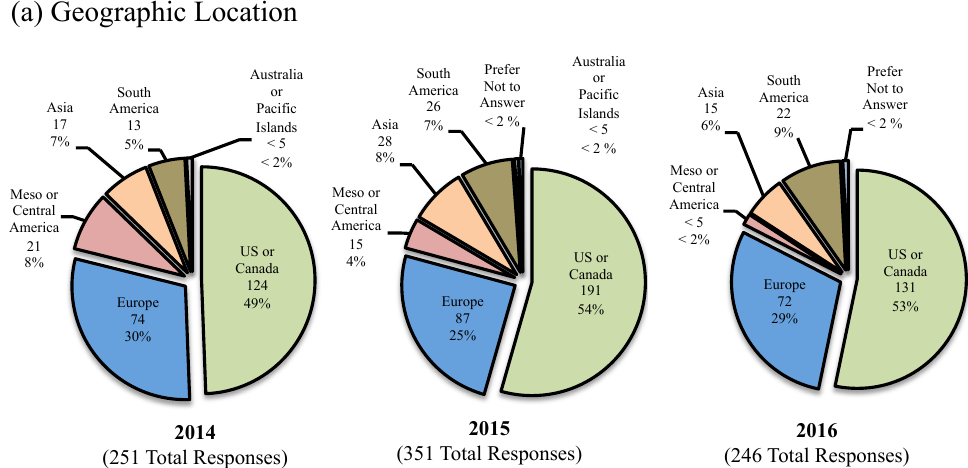} \\
    \includegraphics[width=0.73\textwidth]{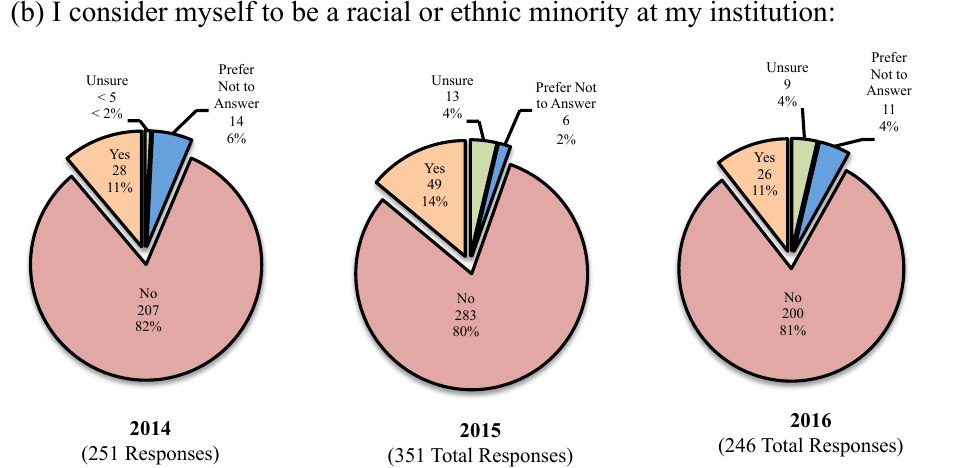} \\
    \includegraphics[width=0.73\textwidth]{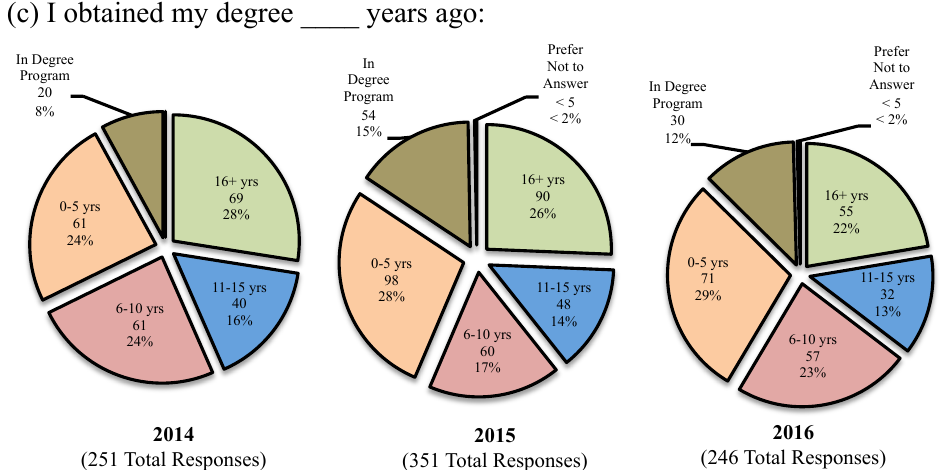} 
    \caption{
    Self-reported demographic characteristics of the SDSS-IV collaboration.
    Each row represents a distinct category of information while each column represents surveys conducted in 2014 (left), 2015 (middle), and 2016 (right). Categories with fewer than 5 responses are shown in the pie charts, with labels ``\textless~5'' and ``($<2$\%)''.
    (a) Distribution of SDSS-IV respondents by geographic location.
    (b) Responses to the question \texttt{I consider myself to be a racial or ethnic minority at my current institution}.
    (c) Years since obtaining highest degree as a proxy for academic age. 
    }
    \label{fig:demographics_prt1}
\end{figure}
\setcounter{figure}{0}
\begin{figure}
    \centering
    \includegraphics[width=0.75\textwidth]{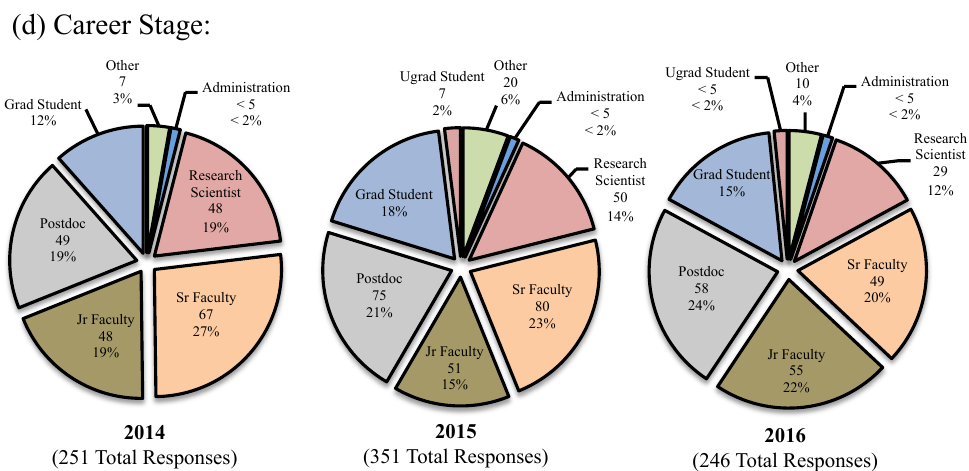} \\
    \includegraphics[width=0.75\textwidth]{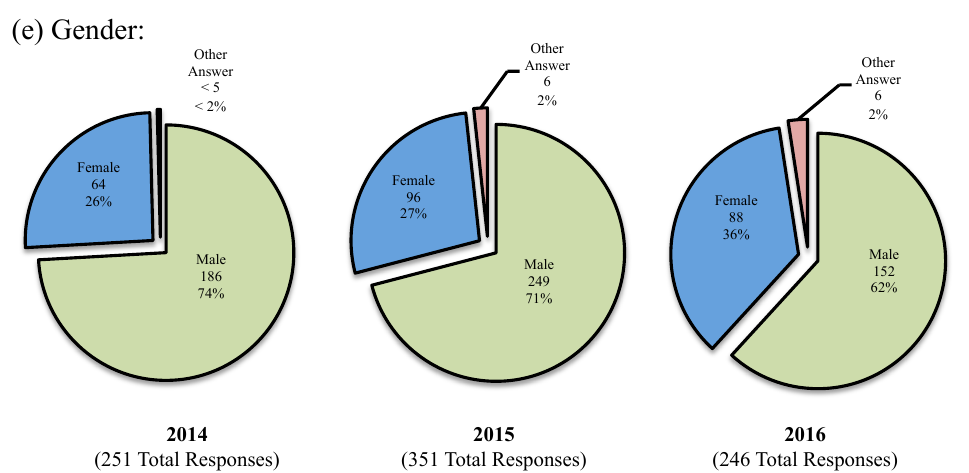} \\
    \includegraphics[width=0.75\textwidth]{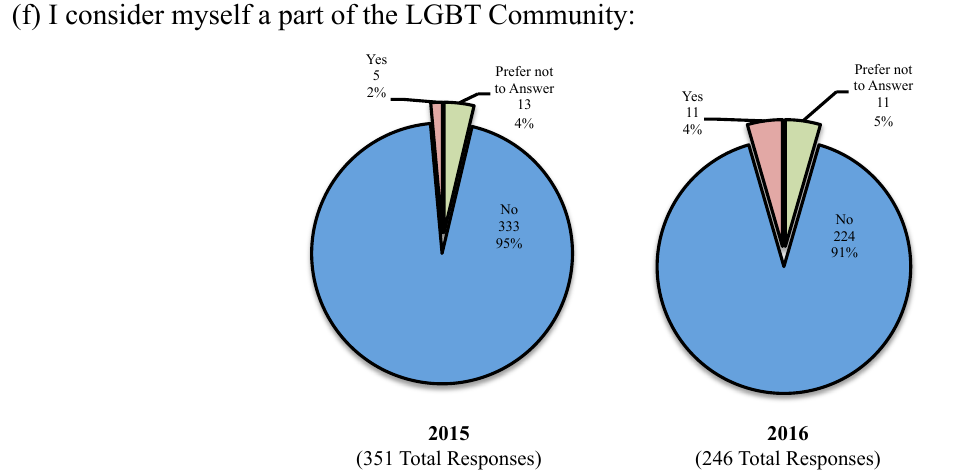} 
    \caption{ 
    {\it -- continued.} 
    (d) Career Stage.
    (e) Gender, note that {\it Other Answer} is an aggregate category with {\it non-binary, prefer not to answer,} and {\it other} to protect the anonymity of those responding in these categories. 
    (f) Responses to the question \texttt{I consider myself part of the LGBT Community.} 
    This question was not included in the 2014 survey. 
    }
    \label{fig:demographics_prt2}
\end{figure}

\section{Overall Demographics of the SDSS-IV Collaboration}\label{sec:overalldemo}

Across the 2014, 2015, and 2016 demographic surveys, we asked the respondents several demographic questions to assess the overall make-up of SDSS. With these data we are able to monitor changes over the three years.  
The answers to a sample of such questions are shown in \autoref{fig:demographics_prt1}. Elements of this figure duplicate those presented in Figure 2 of \cite{lundgren_2015}. 
 
The distribution of geographic location of respondents (\autoref{fig:demographics_prt1}a) remained fairly similar between 2014 and 2016. 
There is a consistent plurality of respondents from the U.S. and Canada ($49-54$\%).
Members working in European countries make up the next largest group ($25-30$\%) of respondents. The number of respondents based in Asia also remained roughly constant ($6-8$\%). 
A notable change across the three years is seen in a decrease of respondents based in Meso or Central America (8\% in 2014 to 2\% in 2016) and an increase in those from South America (5\% in 2014 to 9\% in 2016). This increase could correspond with the development of the Southern hemisphere facility for SDSS-IV at Las Campa\~{n}as Observatory along with the initiation of the Chilean Participation Group (CPG), where the first Memorandum of Understanding agreements (MOUs) were signed in 2015.   

Overall we find that $11-14$\% of SDSS-IV members responding to the survey identify as racial or ethnic minorities where they live or work shown in \autoref{fig:demographics_prt1}b. There are no trends across the three years. As \autoref{fig:demographics_prt1}a demonstrates, SDSS-IV participants are located all over the globe. 
Therefore at some institutions and in some countries, it may be unclear if one is an ethnic or racial minority. Hence, we asked participants to respond if they consider themselves to be a racial or ethnic minority at their current institution. 

\begin{figure}
    \centering
    \includegraphics[width=1.0\textwidth]{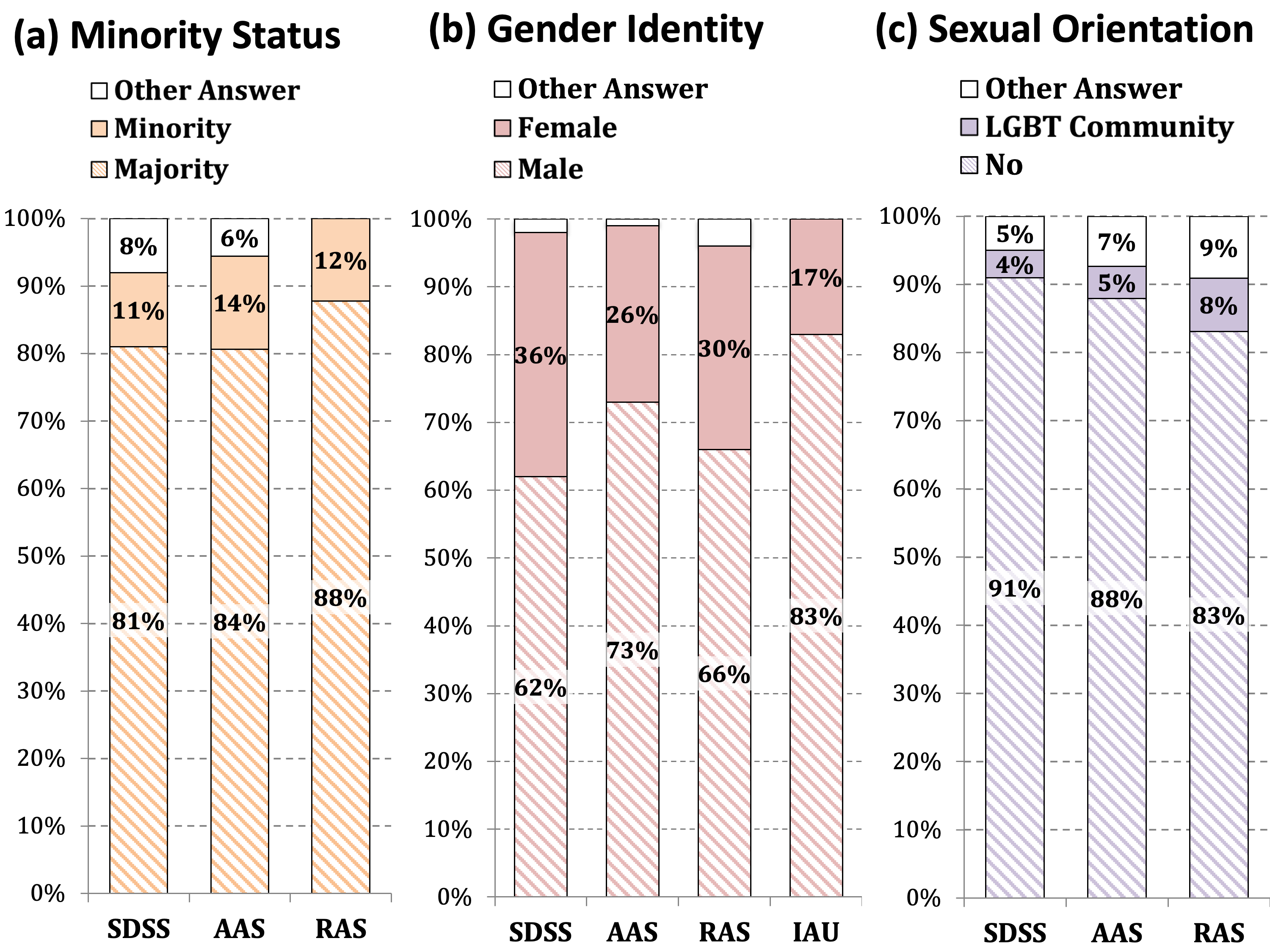} 
     \caption{ Comparison of the SDSS demographic axes of (a) minority status, (b) gender, and (c) sexual orientation between the SDSS-IV survey and values from the AAS Workforce Survey, RAS Survey, and the IAU membership (only available for gender) all from 2016. 
    Results are homogenized into the equivalent SDSS options for AAS and RAS outputs as discussed in the text.}
    \label{fig:demo_comparison}
\end{figure}

We can compare the fraction of respondents that identify as a racial or ethnic minority with the those reported by AAS and RAS to gauge if SDSS is roughly consistent with these larger astronomical societies. The numbers from the 2016 AAS Workforce Survey \citep{pold_2017}\footnote{\url{https://www.aip.org/statistics/reports/astronomy}}, re-binned to mimic our categorization as applicable to the US, are 84\% white (to compare the self-identified ``majority" category in SDSS), 14.4\% Black, Indigenous, and People of Color (BIPOC)\footnote{The categories summed to form BIPOC in 2016 AAS Workforce Study are \citep[see][their Table 25]{pold_2017}: Asian or Asian American (9.1\%), Hispanic or Latino (3.5\%), Black or African American (1.0\%), American Indian or Alaska Native (0.7\%), and Native Hawaiian or other Pacific Islander (0.1\%).} (compared to SDSS self-identified ``minority"), and 5.8\% Other and Prefer not to answer (where Other refers to a category not specifically listed). 
In the 2016 RAS Survey \citep{Massey_2017}, 87.8\% of respondents identified as white (whether British or of another Nationality for comparison with the SDSS self-identified ``majority" category) and 12.2\% identified as BIPOC\footnote{The categories summed to form BIPOC in 2016 RAS Demographic Survey are (Table 16 of the full report available online): Asian:Indian (1\%), Asian:Chinese (1\%),Other Asian (1\%), Mixed: White and Asian (1\%), Mixed: White and Black Caribbean ($<$1\%), Other Mixed Background (1\%).}(compared to SDSS self-identified ``minority").
The fractions from the AAS and RAS are more-or-less consistent with the fractions shown in \autoref{fig:demographics_prt1}b, especially when considering the differences in sampling and nuance in the categorization of minorities in an international context. This comparison is illustrated in \autoref{fig:demo_comparison}a.

We observe an overall increasing fraction of respondents from younger academic age groups (see \autoref{fig:demographics_prt1}c), with 32\% in 2014, 43\% in 2015, and 41\% in 2016 being still in, or within 5 years of, their degree program. This is in contrast with the slight decrease in more senior members (16+ years out from their highest degree) dropping from 28\% to 22\%. 
These trends are also reflected in the \autoref{fig:demographics_prt1}d showing respondents career stage.  
There is an increase in the number of students, postdocs, and junior faculty, and a decrease in the fraction of senior faculty and research scientists responding from 2014 to 2016.  These trends could be due to fluctuations in the response rate or a reflection of SDSS-IV members as a function of career stage. 

Fluctuations in the gender breakdown are shown in \autoref{fig:demographics_prt1}e.  
The fractions of female and male respondents in 2014 and 2015 are consistent, 26 and 27\%. In 2016 the percentage of female respondents increased to 36\%.
There was also a dramatic decrease in the number of male respondents, by 97 people (249 in 2015 to 152 in 2016) compared to the number of female respondents from 96 to 88. 
The increased fraction of female respondents in 2016 could be from a decline in the response rate of male members or an increase in female participation in SDSS. An increase in the fraction of female respondents is in line with the increase of women obtaining astronomy degrees in the  US.\footnote{\url{https://www.aip.org/statistics/data-graphics/percent-bachelors-degrees-and-doctorates-astronomy-earned-women-classes-1}}

We note that our data collection has always included a non-binary gender option. 
In 2014 the option was \texttt{Non-binary/Other} \citep[see][]{lundgren_2015}, but in 2015 and later these categories were made distinct; this is a meaningful change so we note it explicitly \citep[e.g., see discussion in][]{2019BAAS...51g..75R,2021BAAS...53d.442S,2021LPI....52.2306S}. 
In 2016, the option \texttt{prefer not to answer} was also added.  
The number of responses in these categories is small enough that to preserve the anonymity of the respondents we often merge the categories in our figures (e.g., in \autoref{fig:demographics_prt1}e) and analyses.

We can again compare these fractions to other astronomical societies and surveys in 2016, namely AAS Workforce study \citep{pold_2017}, RAS Survey \citep{Massey_2017}, and the IAU membership\footnote{\url{https://www.iau.org/news/announcements/detail/ann16020/}}.  This is shown in \autoref{fig:demo_comparison}b. Overall, SDSS has a slightly higher fraction of women, but is somewhat consistent with these other groups.

In 2015 we added a question about sexuality and gender identify, asking if participants considered themselves part of the LGBT community. We find that across the two years, the number and fraction of people who do consider themselves part of the LGBT community increased from $<2$\% to 4\% (see \autoref{fig:demographics_prt1}f). A similar number of people preferred to not answer this question in both years (13 and 11 people in 2015 and 2016), while 91-95\% of respondents did not consider themselves part of the LGBT community. 
 
National-level astronomical societies also collected data on sexual orientation \citep{Massey_2015,pold_2017}. 
In 2016, 4.7\% of the AAS workforce and 7.8\% of the RAS considered themselves part of the LGBT community, which is similar to 4\% of SDSS members (\autoref{fig:demo_comparison}c).

In the next subsections, we explore the demographics in the following areas:  
    the gender balance in the SDSS-IV collaboration with respect to academic age and career position in \autoref{sec:gender}, and 
    the respondent's educational background in \autoref{sec:edubkgd}.

\subsection{Gender balance breakdown}\label{sec:gender}

\begin{figure}[!ht] 
    \begin{center}
    \includegraphics[width=1.0\textwidth]{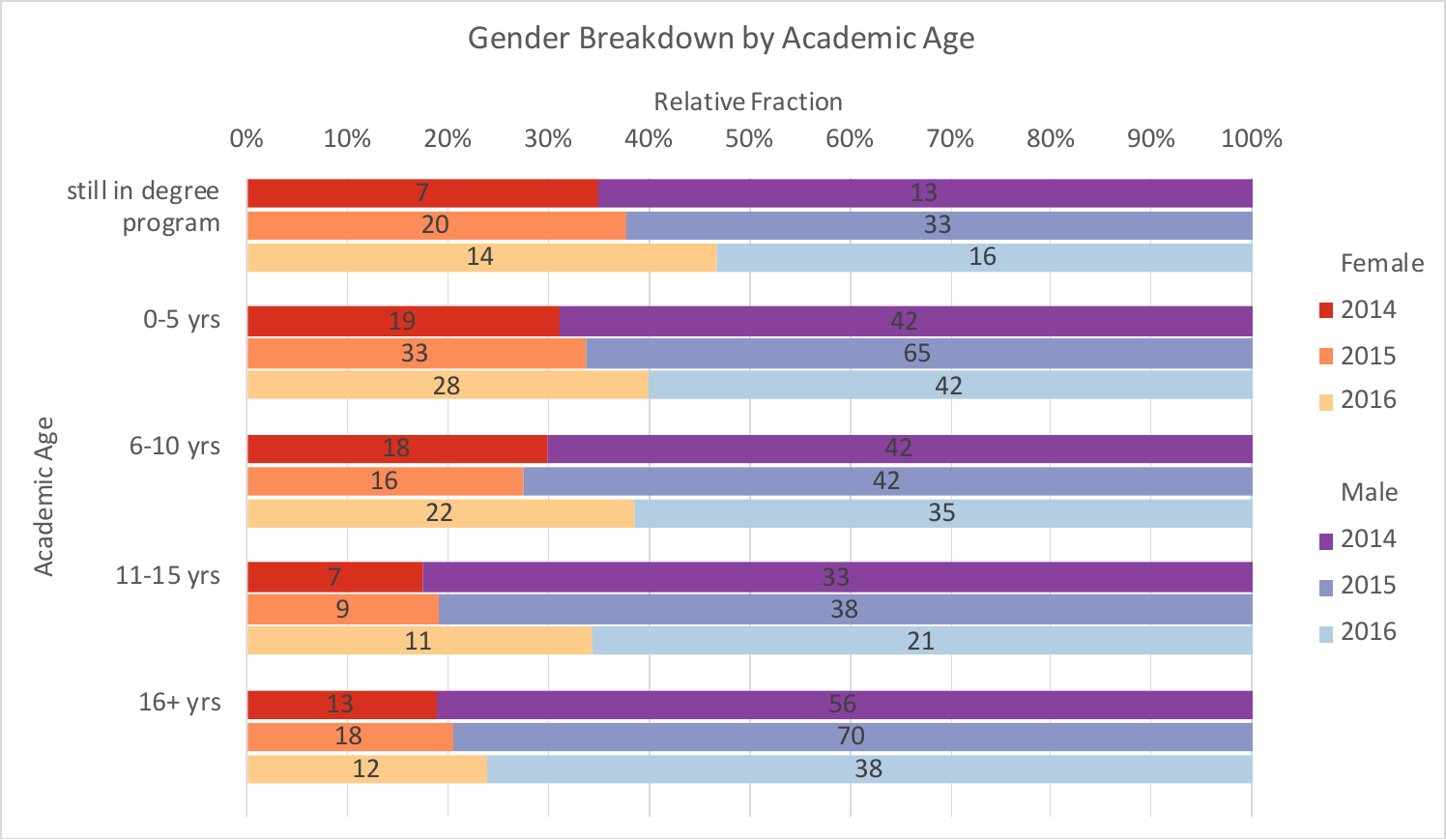} 
    \caption{Gender balance breakdown for the three years of the survey (2014, 2015, and 2016) as a function of academic age, defined as the numbers of years since terminal degree.  
    The relative gender fraction for each year and age are shown, with `Female' on the left and `Male' on the right. 
    The number of respondents in each gender-academic age category is also provided. 
    We were unable to show the non-binary and other responses in this figure due to our commitment to their anonymity.
    The data shown in this figure is available in \autoref{tab:gender}. \label{fig:gender_frac_years}}
    \end{center}
\end{figure}

\begin{table}
    \centering
    \caption{Results from the three survey years,  first broken down by their academic age given as the number of years since they received their highest professional degree (upper) or by career stage (lower) for women (F) and men (M).}
    \label{tab:gender}
    \begin{tabular}{lrrrrrr}
    \hline
     & \multicolumn{2}{c}{2014} & \multicolumn{2}{c}{2015} & \multicolumn{2}{c}{2016} \\
     & F &	 M	& F	&  M & F	&  M \\
    \hline
    \hline
    {\bf Academic Age}:\\
    $<0$ yrs     & 7 & 13 & 20 & 33 & 14 & 16 \\
    $0-5$ yrs      & 19 & 42 & 33 & 65 & 28 & 42 \\
    $6-10$ yrs     & 18 & 42 & 16 & 42 & 22 & 35 \\
    $11-15$ yrs    & 7 & 33 & 9 & 38 & 11 & 21 \\
    16+ yrs      & 13 & 56 & 18 & 70 & 12 & 38\vspace{.5em}\\

    {\bf Career Stage}:\\
    Student & 9 &  20 & 22 & 48 & 17 & 24\\
    Postdoc & 16 &  33 & 27 & 48 & 25 & 33\\
    Junior faculty & 14 & 34 & 17 & 33 & 19 & 36\\
    Senior faculty & 13 & 54 & 15 & 64 & 11 & 33\\
    Research Scientist & 11 & 36 & 9 & 38 & 11 & 18\\
    Administration/Other & $<5$ & 9 & 6 & 18 & 5 & 8 \\
    \hline
    \end{tabular}
\end{table}

In this section we examine the gender balance of the SDSS collaboration as a function of academic age and career position. Since members of the SDSS collaboration have a range of careers and degrees, we consider the academic age to be the time in years since achieving terminal degree;  for most but not all respondents, this degree is a Ph.D. 
While we acknowledge the existence of other genders, here we only compare results based on respondents who self-identify as male or female for anonymity. For the same reason we also do not do this kind of analysis for racial/ethic minority status or being part of the LGBT community.

In \autoref{fig:gender_frac_years} we show the fractions of men and women as a function of academic age based on results from 2014, 2015 and 2016 surveys (the 2014 data was previously presented in Figure 3 of  \citealt{lundgren_2015}). We show the relative fraction of female (red-orange; left) and male (violet-blue; right) respondents for each year of the survey and five categories of academic age. The absolute number of respondents are overlaid and also given in \autoref{tab:gender}. As previously noted, the overall number of respondents peaked in 2015, however all three years present similar fractions by gender with academic age. In general, the fraction of female participants in SDSS-IV decreases with increasing academic age;  from $\sim$35 to 45\% at the student level to $\sim$20 to 25\% in the most experienced age bin.

When we consider the gender balance by career stage instead of academic age, we see a similar trend; i.e. there being fewer female members compared to male members at more senior career stages. These numbers are also provided in \autoref{tab:gender}. Across all years, graduate students, postdocs, and junior faculty are all found to be $30-35$\% women; the fraction decreases for the later career stages, including senior faculty and research scientists, to about 20\%.

As also seen in \autoref{fig:demographics_prt1}e, the overall fraction of female respondents increased each year (from 26\%, 27\%, to 36\%). In 2016, the percentages of female respondents are uniformly higher than the previous year, but the decreasing trend of fraction of respondents who identify as female with academic age remains (\autoref{fig:gender_frac_years}).  This trend of worsening gender balance with academic age is consistent with the commonly referred to phenomenon of the `leaky pipeline', where there are fewer female astronomers farther along the career path \citep[see additional discussion in][]{roy_2020}. 

The leaky pipeline is also seen in the AAS Workforce studies. The 2016 study separated the gender fractions by age, using the birth year 1983 to select those scientists over and under 33 years of age \citep{pold_2017}.  
This roughly corresponds to an age where most scientists have completed a Ph.D. and had some postdoctoral experience. 
Using this split, those born before 1983, the more senior astronomers, were 78\% male, 21\% female, and 1\% prefer not to answer whereas those born after 1983, the more junior astronomers, were 53\% male, 46\% female, and 1\% prefer not to answer. This agrees with the trends seen in SDSS.

The 2018 Global Survey of Mathematical, Natural, and Computing Scientists, designed by the Gender Gap in Science Project and the American Institute of Physics (AIP), collected data from 32,346 scientists in the following eight disciplines: Astronomy, Biology, Chemistry, Computer Science, History of Science, Mathematics, Applied Mathematics, and Physics \citep{roy_2020}. 
This survey aimed at not only describing the gender gap, but in quantifying differences in experiences by women and men at specific developmental periods.
In a multi-variant analysis that accounted for discipline, age, employment sector, geographic region, and level of development, women were: 
    (1) 14 times more likely than men to report having been personally harassed with  $>25$\% of respondents reporting personal harassment at school or work and women more likely to have witnessed sexual harassment,
    (2)  more likely to report discrimination on the basis of gender and less likely to report respectful treatment by co-workers, 
    (3) 1.6 times more likely to have experienced interruptions in their studies, 
    (4) more likely to report less-positive relationships with their doctoral mentors, and
    (5) less-likely to say that everyone is treated fairly. 
\citet{roy_2020} conclude that women and men have very different experiences in their scientific training and work environments. 
Notably, these results held even when controlled for the level of economic or human development and, indeed, often the higher levels of development were correlated with some negative experiences. These findings can help explain the leaky pipeline seen in SDSS and, more generally, in science careers.

\subsection{Educational/Socioeconomic background} \label{sec:edubkgd}
From the 2015 survey onwards, a question was added to assess the educational background of SDSS members. 
This question was added as a way to capture information on socio-economic status, since asking direct questions on the economic and social family background would have been extremely difficult, given the very broad geographical and cultural range of the SDSS collaboration. More specifically, terms such as ``middle class'' or ``working class'' are relative, subjective and strongly dependent on geographical and cultural backgrounds \citep[e.g.,][]{BestPractices_SocialClass}. 
Thus, the highest education-level attained for the parents/guardians is an easily accessible proxy for socio-economic status.
There are also well documented correlations between parental educational achievement and income \citep[for example in the US see, ][]{Census:Attainment,Census:Attainment2} and between  parental educational achievement and access to educational opportunities and quality of instruction \citep[e.g., ][]{doi:10.3102/01623737020004253,doi:10.1177/0003122414546931}.
In the 2015 and 2016 surveys (see \autoref{app:survey}), question 16 asks \texttt{what is the highest level of education achieved by the parent/guardian}.  
Respondents were instructed to select the highest level of education of any parent or guardian, with 
 possible answers ranging from \texttt{Did not complete primary school} to \texttt{PhD/Doctoral degree} (see \autoref{fig:2016_survey_questions}). 

\begin{figure*}[h] 
\begin{center}
    \includegraphics[width=0.3\textwidth]{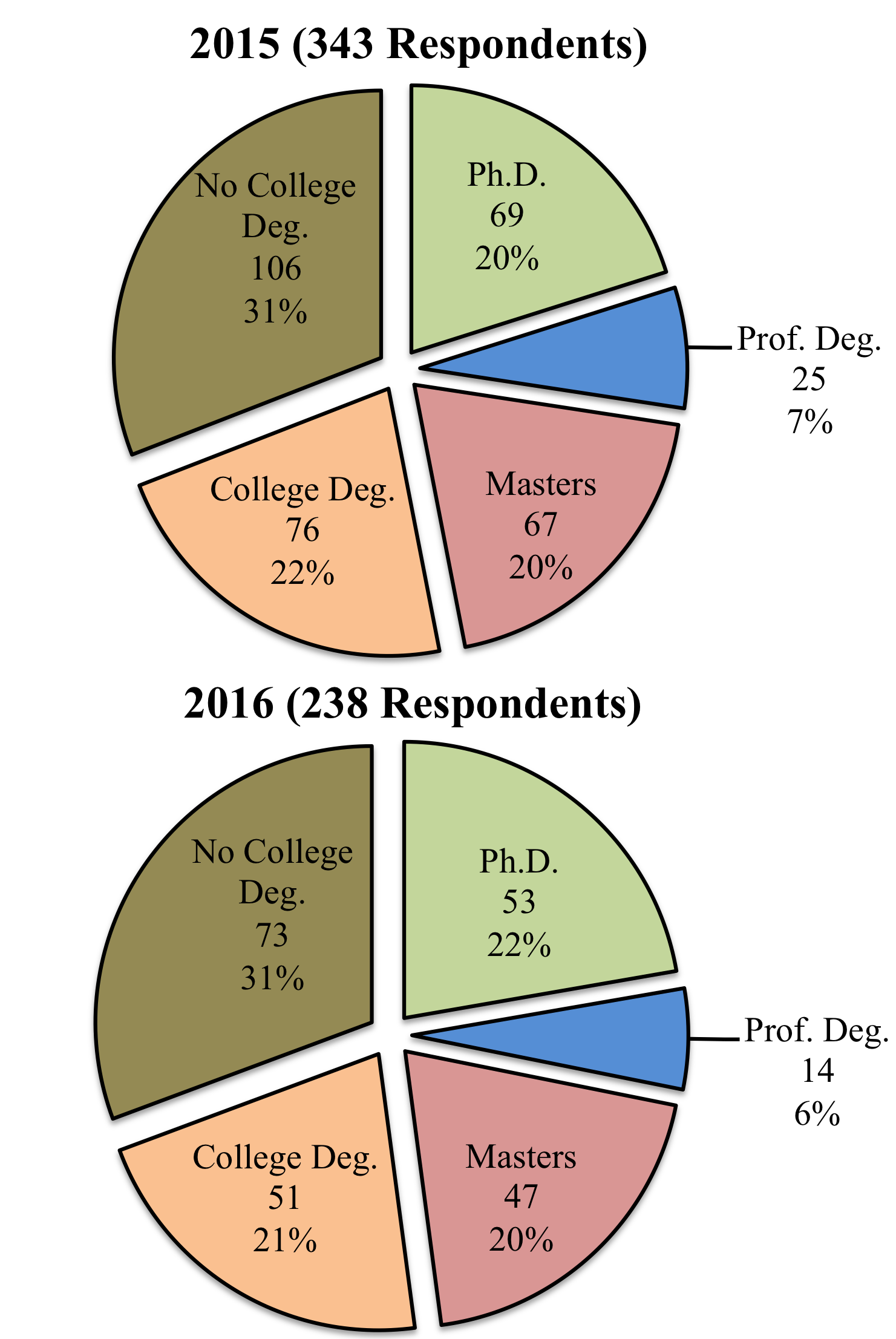}
    \includegraphics[width=0.6\textwidth]{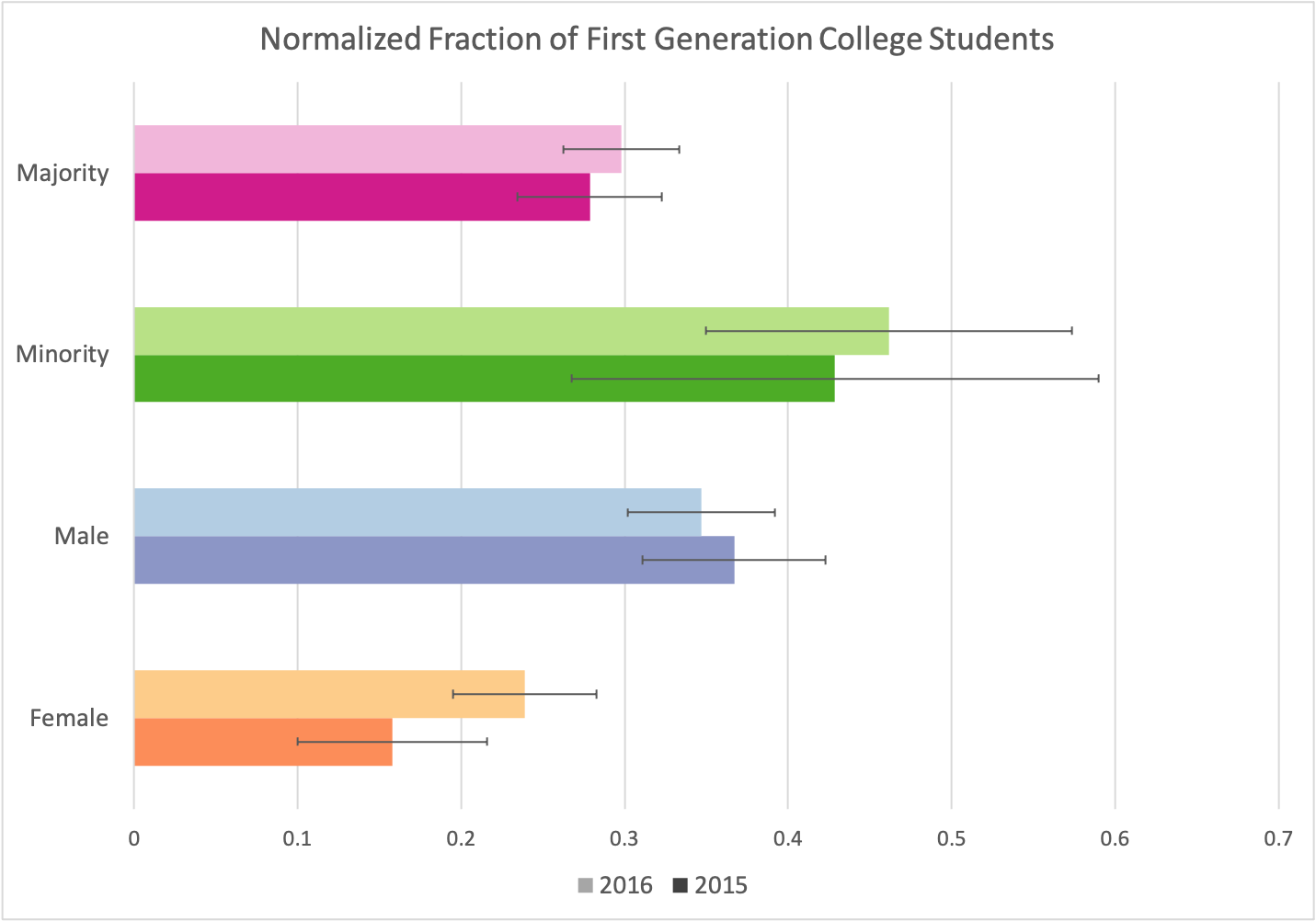}
    \caption{The two pie charts on the left show the distribution of responses to the question \texttt{what is the highest level of education achieved by the parent/guardian}, with ``no college degree'' being a combination of \texttt{did not complete primary school}, \texttt{primary school}, \texttt{high school}, and \texttt{some college}.  The left top pie chart is from 2015 and bottom left from 2016. The right side presents a comparison of the normalized fraction and errors of first generation college students (FGCS) for demographic subsets in SDSS from the 2015 (darker color) and 2016 (lighter color) surveys.
    The top two sets compare the normalized fractions of FGCS drawn from a majority (top, pink) or minority (2nd from top, green) racial/ethnic group. 
    The bottom two sets compare the normalized fractions of FCGS drawn from male (2nd from bottom, blue) and female (bottom, orange) respondents.
    The normalized fractions are statistically consistent over the two surveys.
    We see at least a 1-$\sigma$ difference between the normalized fraction of FGCS in the majority to minority comparison, with minorities having a higher fraction and also between the male to female comparison, with males having a larger fraction. 
   }\label{fig:edu_gender}
\end{center}
\end{figure*}

The left panels of \autoref{fig:edu_gender} show the distribution of responses in 2015 (top) and 2016 (bottom) to the survey question. We find that about 69\% of the respondents' parents/guardians have at least an undergraduate-level college degree. 
Although there are variations (by academic age and country), this fraction is significantly higher than the average for the general population in any country hosting an SDSS institution.
For example, among the 38 countries in the Organisation for Economic Cooperation and Development (OECD; including the US, UK, Canada, Chile, Germany, Mexico),  39\% of adults have at least undergraduate-level degrees, with a variation of 45\% in the youngest age bin to 27\% in the oldest age bin \citep{oced_2021}.

The proportion of SDSS-IV members whose parents/guardians have a masters degree is 20\% in both years; coupled with professional degrees, this fraction is 27\% and 26\% with degrees beyond a college degree. The OECD-average is 14\% of the general populations have masters degrees, with the UK and US at 13\% and 12\%, respectively. Thus, SDSS members are 1.4 times more likely to have parents with a masters or professional degree than the average of the general population.
Most strikingly, the proportion of SDSS-IV members whose parents/guardians have a doctoral degree is 20\% and 22\% in 2015 and 2016, respectively, while the OECD-average is that 1\% of adults hold doctoral degrees, with the UK and the US at 2\% \citep{oced_2021}.
Thus, respondents in SDSS are $\sim10$ times more likely than the average of the general population to have a parent with a doctoral degree.

Put together, at least one parent of 48\% (47\%) of respondents in SDSS-IV for 2015 (2016) have educational attainment beyond a college degree, and 69\% in both years having a parent with any college degree. In contrast, the OECD averages are 15\% of the general population with masters or doctoral degrees and 33\% having any college degree \citep{oced_2021}.  
Thus, SDSS-IV respondents are nearly twice as likely to have a parent with a college degree, and thrice as likely to have a parent with an advanced degree compared to OECD country averages. 

We searched for similar statistics obtained by Astronomical or Physical Societies and found that data on this particular demographic axis is not available and, thus, we are unable to evaluate if the fractions in SDSS-IV are consistent with the general trend in the Astronomy and Physics workforce, but such data does exist for general academic faculty \citep[e.g., ][]{morgan_clauset_larremore_laberge_galesic_2021}. 
More specifically, in their study of eight academic disciplines, \citet{morgan_clauset_larremore_laberge_galesic_2021} find that academic faculty are 25-times more likely to have a parent with a Ph.D. than the general population and about two-times more likely than other Ph.D.-holding individuals.
This general trend seems to hold for SDSS respondents in 2015 and 2016. 
Having well educated/high socio-economic status parents seems to make it more likely you will become an SDSS-IV collaborator. 

\subsubsection{Gender, and Socioeconomic Status/Educational Background}
\label{sec:edugender}
There are sufficient number of respondents to divide the educational attainment by gender. We find that female survey respondents are more likely to have a parent with a PhD (a 1-$\sigma$ increase), while male survey respondents are more likely to have a parent with a Masters degree. An early analysis of the results on gender and educational background were previously commented upon in an opinion piece \citep{lucatello_2017}, where it was speculated that this was evidence that women need higher educational background/socioeconomic status to persist in science careers than men. 

\subsubsection{Analysis of First Generation College Students in SDSS-IV}
\label{sec:fgli}

We re-organize the data to collect all first generation college students (FGCS) into a single group, defined as those not having either parent with a college degree (i.e. respondents that selected \texttt{did not complete primary school}, \texttt{primary school}, \texttt{high school}, or \texttt{some college}). This makes up 31\% of SDSS respondents in 2015 and 2016 (\autoref{fig:edu_gender}). 
On the right side of \autoref{fig:edu_gender}, the normalized fractions of FGCS are compared for both 2015 and 2016 respondents (light and dark tones, respectively) using two demographic axes: 
    (i) membership in a minority or majority racial group (see Section 3 for a description of the reason for this language; \autoref{fig:demographics_prt1}b) and 
    (ii) binary gender (non-binary answers are not shown to maintain anonymity; \autoref{fig:demographics_prt1}e). Along  each demographic axis, the normalized fraction gives the number of FGCS out of the total number of respondents in that category.

We find that between $40-50$\% of respondents self-identifying as in a minority racial group are FGCS, a notable increase over the $25-30$\% of respondents in the majority racial group (see the top two bars in \autoref{fig:edu_gender} -- where the majority is shown at the top (pink bars) and minority second from top (green bars)).
Low numbers of minority respondents result in large uncertainties in the fraction of FGCS, so while a notable difference, this  comparison is only minimally statistically different.

We find that the normalized fraction of FGCS among male respondents is $\sim$35\% both years, while for women it is just 15\% (2015) or 25\% (2016). This is shown visually in the lower two sets of bars of \autoref{fig:edu_gender}, which shows the results for male respondents (second from bottom, blue) and female respondents (bottom, orange). The difference between these fractions is more than 1$\sigma$. Female FGCS sit at the intersection of two under-represented groups in professional-STEM, being both first generation college students (FGCS) and female.
Thus, this notable under-representation at the level of SDSS-IV participation, while disappointing is expected, as we briefly summarize. 

FGCS face a number of barriers related to how socio-economic class imparts ``cultural capital'' in terms of how to leverage experiences in higher education.
Particular challenges for FGCS are often encapsulated with the term ``hidden curriculum'' that FGCS have to navigate. 
Components of the ``hidden curriculum'' vary from aspects that impact individual course performance, for example, knowing that ``office hours'' are open hours for students to engage with instructors (and not quiet time when the professor should not be disturbed), to those that have impacts on career progression or competition for post-graduate experiences, for example, seeking out strong inter-personal relationships with professors that result in stronger letters of recommendation or nominations for prizes (among others). 
Students with prior exposure to the ``hidden curriculum'' perform better in college and are more likely to persist \citep{Anyon_ClassHiddenCurriculum,doi:10.1080/00221546.2016.11780889}. 
In particular, studies have indicated that socio-economically disadvantaged students attending elite preparatory schools are exposed to ``elite institutional cultures'' and this, to some level, ameliorates the impact \citep{Jack+2019,doi:10.1177/0038040715614913}.

Studies have shown that the development of a scientific identity is related to being more resilient in STEM progression and, thus, a component to persistence in STEM in the face of personal or institutional barriers; this is particularly true for women and women of color \citep{Carlone_Identity,Hazari_Gender,Ong_Counterspaces}. 
The impact of the cultural transition is more than just perceptual and is accompanied by  physiological impacts measured by increased levels of stress hormone that in turn impact one's ability to perform at their best capability \citep{STEPHENS_Cortisol}.
Active encouragement from faculty is particularly important to developing a STEM identity \citep{Carlone_Identity}.
That FGCS may require additional institutional and structural support is well established \citep{rondini2018clearing}, and that those facing other inclusion intersections, such as gender or race, will have additional barriers. 
However, in the case of SDSS our membership is drawn from an international perspective and thus additional concerns may apply in differing educational environments \citep{thomas2006first} or when facing additional language and/or cultural barriers when studying outside their native country.

 \section{Demographics of the SDSS-IV Collaboration Leadership}\label{sec:leadership}

Here, we present the leadership demographics of SDSS-IV from the three years of demographic survey data.  
We have examined leadership data as a function of gender, career position/age, collaboration structure and social status; we provide these data in \autoref{tab:lead_gend_age} and \autoref{tab:lead_other}. 
Due to low numbers on most of these axes, and many possible unknown factors, we focus primarily on gender representation in leadership roles.

In the next sections, we define ``leadership'' as it was used in the survey (\autoref{sec:leadership_def}) and  describe how members became leaders in SDSS-IV (\autoref{sec:leadership_paths}). We give a general breakdown of the results of number of leaders by gender (\autoref{sec:leadership_break}), considering 
gender balance and recognition of leadership (\autoref{sec:leadership_rec}), and funding of leadership roles (\autoref{sec:leadership_funds}). We go on to consider the impact of gender on the path to leadership in \autoref{sec:leadership_adv}. 

Other astronomical organizations also study the demographics of leadership. Mission Principal Investigators (PIs) can be considered leaders in the field as it requires substantial knowledge, technical expertise in hardware and operations, as well as team management skills.  The NASA Science Mission Directorate (SMD) recently commissioned a study to examine the diversity pool of proposal leaders of PI-led space or Earth science missions between 2010-2019.  Of the 101 submitted proposals in the field of Astrophysics, only 7\% were female and 93\% male, see \citet[][their Table 3.2]{natacademy2022}, with no female-PI proposals funded.

A study of gender demographics of HST proposals 
\citet{Reid2014_HSTGender} found that across 10 years, the fraction of female-PIs of submitted proposals was 21\%, compared to male-PIs at 79\%.  
The success rate of female-led proposals was systematically lower at 19\%, compared to males with 23\%.  
Interestingly, the first year after switching to a dual-blind submission/review process \citep{Johnson_2020,2021BAAS...53d.010A}
the success rate equilibrated to 8.7\% and 8\% for women and men, respectively.

\subsection{Definition of Leadership in SDSS-IV}\label{sec:leadership_def}

Our analysis uses self-reported leadership status, which means a positive \textbf{yes} answer to the question ``\texttt{I currently consider myself to be in a leadership or decision-making role [official or unofficial] within SDSS-IV}''. 
Feedback on the 2015 Demographic Survey indicated that the definition of leadership was unclear and potentially confusing.  
Therefore in the 2016 survey, the following definition, provided in the survey introduction, was copied into the first question about leadership in SDSS-IV:
\begin{quote}
``\texttt{For the purposes of this survey, we intend "leadership" and "leadership role" to refer to any role whose tasks or responsibilities include making decisions that affect other people and the survey, organizing regular project discussions or meetings, professional mentoring, or influencing/directing others in their tasks.}'' 
\end{quote} 
\noindent The bottom-left panel of \autoref{fig:2016_survey_questions} in  \autoref{app:survey} provides an example of how this looked in the survey.  Leaders were asked a number of follow-up questions to characterize their leadership role (see \autoref{app:survey} \autoref{fig:2016_survey_questions} for the formal wording), that included whether or not their role was formally recognized by the survey (hereafter, \textbf{recognized leaders}), if their role was funded or partially funded by SDSS-IV (hereafter, \textbf{funded leaders}), and a question regarding how they came into their leadership role (hereafter, the \textbf{path to leadership}).  \autoref{tab:leadership} displays a summary of the results for all responses, leaders, recognized leaders, and funded leaders split by gender, female (F) or male (M) (other responses, including non-binary genders, are not shown for anonymity).

Roughly one third of all respondents self-identify as being a SDSS-IV leader per the definition above, with the percentages being 35\%, 32\%, and 38\% for 2014, 2015, and 2016, respectively. 
The change in question wording between the 2015 and 2016 surveys may have had a minor effect in the way people responded, because relatively more people self-identified as a leader in 2016. 

\subsection{How Members Became Leaders in SDSS-IV}\label{sec:leadership_paths}

There were a variety of ways that members become leaders in SDSS-IV as it pertains to formally recognised and/or funded roles within the SDSS collaboration. 
Three common paths to leadership are as follows: 
\begin{enumerate}
    \item \textbf{Contractual Leadership:} Some of the Memorandums of Understanding (MOUs) between SDSS-IV and partner institutions include explicit agreements to provide effort toward certain tasks in the form of a recognized leadership role (e.g., filling a position on an organizational chart).  
    Often, but not always, the MOUs are constructed in this manner precisely because the person currently serving in that role is employed at that specific institution prior to the signing of the MOU. 
    \item \textbf{Open Calls:} 
    SDSS-IV leadership developed a policy to implement open calls for applications to fill future position vacancies; the resulting policy document is reproduced in \autoref{sec:fillingroles}. 
    After this point, all openings for formally named positions in the Organization chart were supposed to be advertised to SDSS-IV members following the guidelines in this document. 
    For example, the role of Spokesperson in SDSS-IV was an open call for nominations (including self-nominations), followed by an election (with all SDSS-IV members being eligible to vote).  Another example are the roles of Working Group Chairs within the constituent surveys that were advertised in the survey mailing list with a description of expected work and requirement of submission of a statement of interest; the leader(s) was then selected from this applicant pool to fill the position. As described in \autoref{sec:fillingroles}, specific candidates were often recruited into a position and, based on the Sloan Foundation recommendations, female candidates were specifically encouraged.
    \item \textbf{Task-Based:} More informal, but still ``recognized" leadership positions, at times, have also emerged organically within SDSS-IV operations. 
    Specific scientists who volunteer to take on responsibilities may end up with named positions on an Organizational chart defined around work they were already informally leading, but their leadership and responsibilities became formalized to recognize its importance. 
\end{enumerate} 

The definition of leadership used in the Demographics survey also includes other leadership roles beyond those explicitly included in the descriptions above.
Thus, we also ask about self-identified leadership that is not formally recognized as a leadership position in the Organization Chart. 
Such roles may include but are not limited to: 
    (1) informal mentoring of junior members, 
    (2) work on specific initiatives not directly funded by SDSS-IV, 
    (3) organizing smaller working groups on specific science topics, 
    (4) service on committees like Education and Public Outreach (E/PO) and the Committee on INclusivenss in SDSS (COINS), and, 
    (5) voluntary participation in infrastructure tasks such as documentation efforts around data releases or construction of important value-added data products.

The Demographic Survey captured paths to leadership with the survey question ``\texttt{I ended up in this role via:}'' (see \autoref{app:survey} \autoref{fig:2016_survey_questions}), which had the following possible options:
\begin{quote}
\begin{tight_enumerate}
    \item \texttt{Was explicitly encouraged to apply for position}
    \item \texttt{Responded to an open call for applicants}
    \item \texttt{Was asked to fill position without a formal application}
    \item \texttt{Position was defined around work I was already doing}
    \item \texttt{Was asked by others to take on tasks that ended up defining this role}
    \item \texttt{Prefer not to answer}
    \item \texttt{I don't consider myself in a leadership role}
    \item \texttt{Other}
\end{tight_enumerate}
\end{quote}
Respondents could check as many that applied to their leadership roles(s). 

 \begin{figure}[ht]
    \centering
    \includegraphics[width=1.0\textwidth]{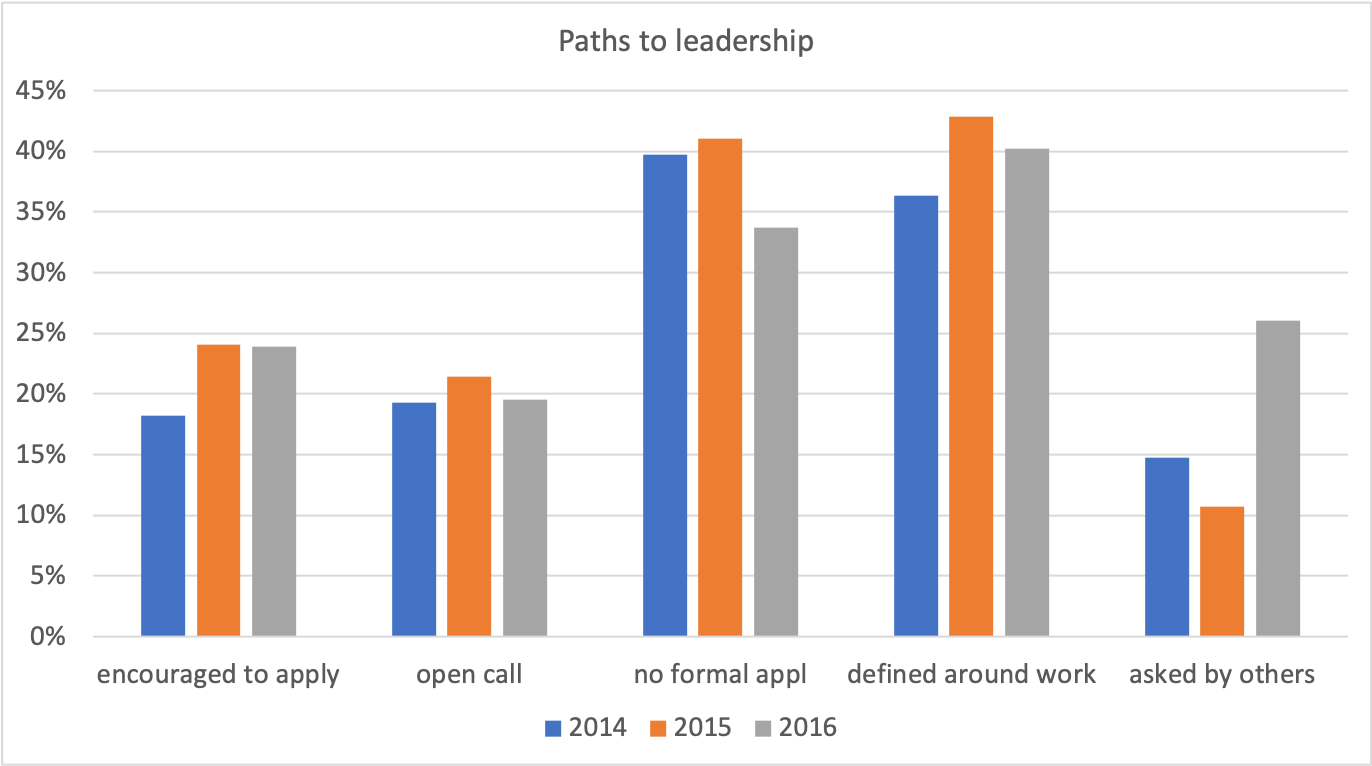}
       \caption{Comparison of five paths to leadership in SDSS for self-identified leaders for the three survey years 2014 (blue), 2015 (orange), and 2016 (grey). Respondents were able to select multiple options for this question, with the highest individual rates for \texttt{no formal application} and \texttt{defined around work I was already doing}.   
       \label{fig:leadership_paths}}
\end{figure}
 
Comparing the open call fraction across the three years of the survey also indicates there was no significant change, despite the implementation of a policy to advertise roles (see \autoref{sec:fillingroles}). 
Broadly, these results highlight that even with an explicit policy, the overall number of roles filled through the open call path is small compared to those filled by contractual and task-related paths. 
As a result, we may not see gains in the overall demographics of leaders that would be anticipated by having open calls.

\begin{table}[h]
\centering
    \caption{Gender breakdown of self-identified leaders for the three years of the survey.  
    \label{tab:leadership}
    }
\begin{tabular}{lrrrrrr}
    \hline
     & \multicolumn{2}{c}{2014} & \multicolumn{2}{c}{2015} & \multicolumn{2}{c}{2016} \\
     & F &	M 	& F 	&M  &	F 	& M\\
     \hline
     \hline
    All responses &	64	& 186 &	96 &	249	& 88 &	152\\
    All leaders & 20  & 68  & 24  & 85  & 27  &	62  \\
    Recognised leaders  & 16  & 56  & 16  & 66  & 15  &	53  \\
    Funded leadership roles & $<5$ & 21 & $<5$ & 27 & 6 & 22 \\
    \hline
\end{tabular}
\end{table}

\subsection{Gender Breakdown of Leaders in SDSS-IV}\label{sec:leadership_break}

The total number of survey respondents who are male is higher than those who are female (see \autoref{fig:demographics_prt1}e), and we also find that the the number of male leaders is significantly larger than the number of female leaders, see in \autoref{tab:leadership}. 
We will explore leadership fractions normalized by the total number of male and female respondents rather than the absolute number of leaders. 
The normalized fraction gives the number of female (male) leaders out of the number of female (male) respondents and may reveal gender-related bias in self-reported leadership.

\begin{figure}[h]
    \centering
    \includegraphics[width=1.0\textwidth]{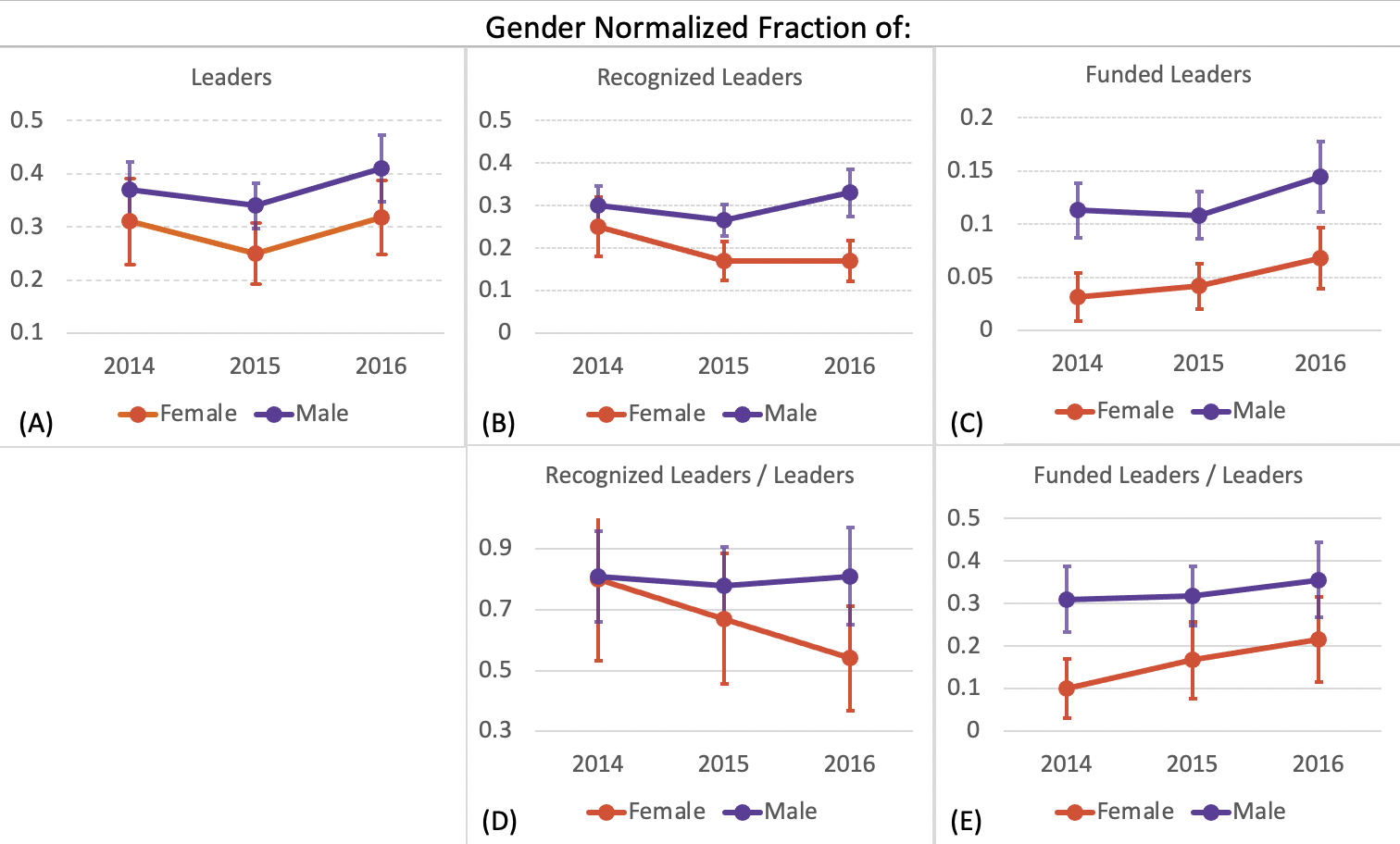}
       \caption{
       Gender breakdown of people in self-identified, recognized and funded leadership positions across the three years of the survey.  Shown here are the gender normalized fractions: for leaders as (A) the number of female (male) leaders over the number of female (male) respondents, recognized leaders (B) recognized female (male) leaders over female (male) respondents, funded leaders (C) funded female (male) leaders over female (male) respondents, (D) female (male) recognized leaders over female (male) leaders, and as (E) funded female (male) leaders over female (male) leaders.
       Note that in 2016 the question on leadership was further clarified in the text with the definition of ``self-identified leadership" copied from the introduction part of the survey to the question description. 
       The data which created these plots is shown in Table \ref{tab:leadership}.\label{fig:leadership_rec_adv}}
\end{figure}

\autoref{fig:leadership_rec_adv}a compares the gender-normalized fraction of leaders for male (purple) and female (red) respondents for 2014, 2015, and 2016.  The uncertainties shown are propagated Poisson errors for the given fraction.
A higher fraction of male respondents are leaders than female respondents, though the uncertainties overlap. 
Between 25 to 31\% of female respondents self-identify as leaders (31\%, 25\%, 31\% in 2014, 2015, 2016), while for men it is 37\%, 34\%, and 41\% for 2014, 2015, and 2016. Overall, male respondents in SDSS-IV are slightly more likely to self-identify as having a leadership role.  
This may be due to many factors, such as there being more male leaders overall, that male respondents are more likely to self-identify as compared to female respondents, or that female collaboration members who are not leaders are more likely to respond to a survey than male collaboration members who are not leaders.

\subsection{Gender Balance and Recognition of Leadership}\label{sec:leadership_rec}
 
In this section we consider those in leadership positions who feel their position is officially recognized within SDSS-IV. We consider a \textbf{recognized} leader as those who replied \texttt{yes} to the prompt \texttt{The primary role in which I lead others is officially recognized within the survey (e.g., a named position on an org chart)}. The number of recognized leaders is presented in \autoref{tab:leadership}.  

\autoref{fig:leadership_rec_adv}b shows the number of female (male) recognized leaders over the total number of female (male) respondents (the normalized gender fraction as in \autoref{fig:leadership_rec_adv}a).  
As seen in \autoref{fig:leadership_rec_adv}a, there are fewer female recognized leaders compared to male recognized leaders for all three years, with the separation growing larger with time and becoming larger than the statistical errors in 2016.  
  
To understand if this trend is mostly a reflection of the difference seen in the rate of leaders in \autoref{fig:leadership_rec_adv}a, we plot in \autoref{fig:leadership_rec_adv}d the number of female (male) recognized leaders over the number of female (male) leaders.  
The fraction of male recognized leaders stays consistent for all the three years.  
For women, while the fraction of recognized leaders was the same as men in 2014, by 2016 it has decreased relative to male leaders by more than one standard deviation.  This decrease of female recognized leaders may be correlated with the increase of younger female respondents in 2016.  
However, these numbers are small and the uncertainties can be large, so it is difficult to make any conclusions. 
  
Part of the formal SDSS inclusion policies (see \autoref{app:current}) is to recognize leaders on an easily accessible and well-maintained organizational chart called \textbf{key personnel}; such a process can also result in giving positions formal names. We can compare the fraction of recognized leaders from the survey with those on the organizational chart.
To examine the key personnel, we tabulate the percentages of male and female members in \autoref{tab:orgchart} (note: this is based on a non-ideal way of identifying gender based on presentation and is not self-identified).
This organizational chart is top-down with project-wide roles in the ``top'' spots and individual-survey roles in the lower spots (e.g., APOGEE, eBOSS, or MaNGA specific roles); thus we can further break down the organizational chart with the top-$N$ positions (exact number can vary with time) as being those with the highest importance or greatest responsibility within the survey. 
We find a severe lack of female leaders in the highest $N$ positions, but the fraction does slightly increase, 13\% to 25\%, from 2014 to 2016.  
For 2016 we also compare the gender breakdown of all the key personal (about 160 positions) and find about 23\% are female and 77\% are male, similar to the top $N$ position fractions for that year.  Comparatively the 2022 leadership team at the Space Telescope Science Institute (STScI) is 48\% female, and 52\% male.  While the comparison year is not the same, it shows that it is possible to have an astronomical institute with roughly 50\% female leaders. 

\begin{table}[h]
\centering
\caption{Percentage gender breakdown of the SDSS org chart by year.  ``Top N" indicates the numbers in the top level of the org chart ($N_{pos}$ as listed). "Key Personnel" indicates the percentage breakdown of members of SDSS in any named position on the org chart (N$\sim$160). While this presents F/M binary genders we acknowledge non-binary genders exist, but we are not aware that they are represented in this group.
\label{tab:orgchart}}
\begin{tabular}{lrrrrrr}
\hline
 & \multicolumn{2}{c}{2014} & \multicolumn{2}{c}{2015} & \multicolumn{2}{c}{2016} \\
 N$_{pos}$ & \multicolumn{2}{c}{30} & \multicolumn{2}{c}{34} & \multicolumn{2}{c}{39} \\
 Gender & F &	M 	& F 	&M  &	F 	& M\\
 \hline
 \hline
Top N &	13\%	& 87\% &	19\% &	81\%	& 25\% &	75\%\\
Key Personnel &		&  &	 &		& 23\% &	77\%\\
\hline
\end{tabular}
\end{table}

Contrasting the increase in ``officially'' recognized leadership roles from the organizational chart (Top $N$) for women over time (\autoref{tab:orgchart}), with the decrease in self-perceived recognized leadership roles for females from the survey \autoref{fig:leadership_rec_adv}b and \ref{fig:leadership_rec_adv}d), could suggest that while female leaders have grown at the top level positions, women in the collaboration at large feel more unseen and unrecognized in SDSS.  While the numbers are small, and conclusions are difficult to draw, it suggests a disconnect between the SDSS leadership team and its collaboration members.

\subsection{Gender Balance and Funded Leaders} \label{sec:leadership_funds}

In this section we examine the gender balance with respect to those in leadership positions who receive financial compensation in exchange for their leadership role.  We consider a \textbf{funded} leader as those who replied \texttt{yes} to the prompt \texttt{This role is an SDSS-funded position (full or partial salary)}. The number of funded leaders is presented in \autoref{tab:leadership}.  The number of funded leaders is significantly lower than the numbers of all leaders or those in recognized leadership roles across all three years.   

The right-most column of \autoref{fig:leadership_rec_adv} focuses on the gender breakdown of funded leaders.  As was done with recognized leaders, \autoref{fig:leadership_rec_adv}c shows the number of female (male) funded leaders over the total number of female (male) respondents, while \autoref{fig:leadership_rec_adv}e shows the number of female (male) funded leaders over the total number of female (male) leaders.  

In \autoref{fig:leadership_rec_adv}c, we find that funded leaders make up a small fraction of overall people in the survey, at most 15\%.  Additionally the fraction of male and female funded leaders differ significantly across all three years, with women occupying a much lower fraction ($\lesssim$7\%) of funded leadership positions than men.  In \autoref{fig:leadership_rec_adv}e, we also find a significant difference between male and female funded leaders, with funded male leaders staying consistent at 30-35\%, but funded female leaders making up $\sim$10 to 21\% in 2014 to 2016.

\subsection{Gendered Differences in How Members Became Leaders in SDSS-IV}\label{sec:leadership_adv}

Having looked at the gender balance of leaders in various types of leadership positions, we now revisit the paths to leadership in a gendered context using the response breakdown of leadership paths shown in \autoref{fig:leadership_paths}.  For each of the 5 itemized categorical paths, we compute the gender normalized fraction of leaders, similar to \autoref{fig:leadership_rec_adv}.  

\begin{figure}[h]
    \centering
    \includegraphics[width=0.5\textwidth]{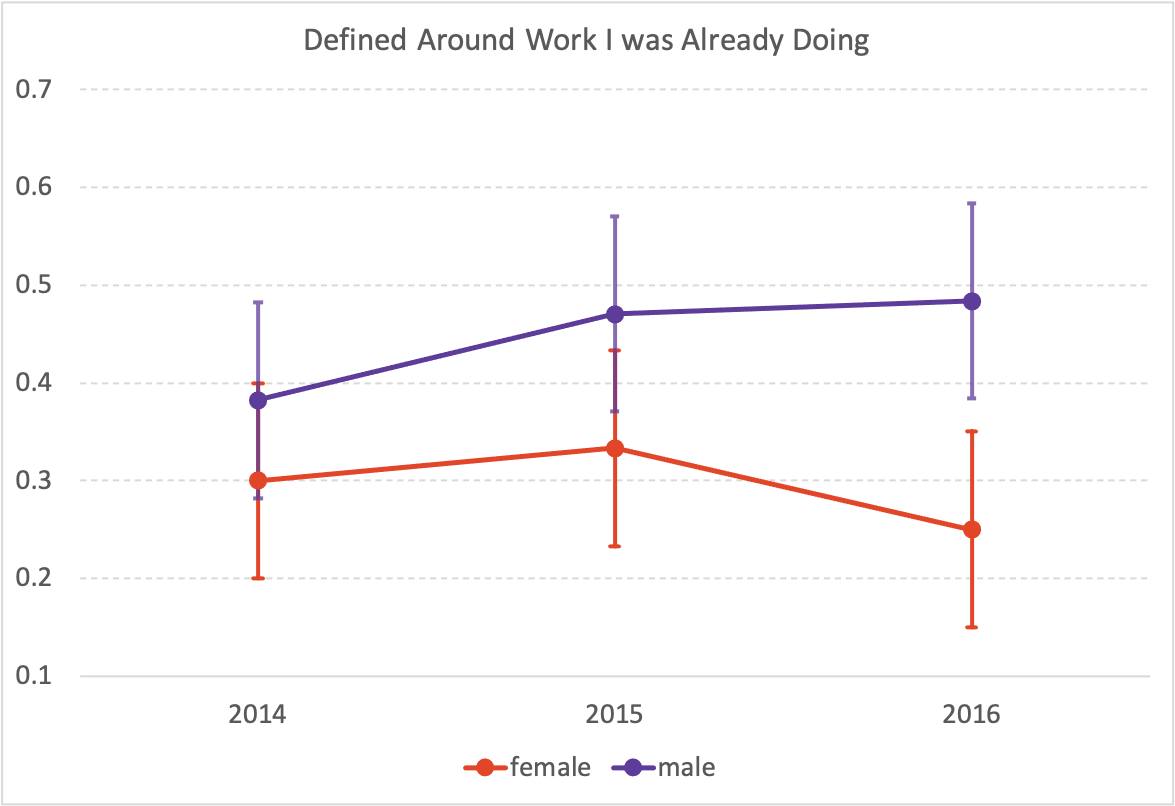}
       \caption{Gender-normalized fraction of the leadership path category: \texttt{Defined Around Work I was Already Doing} for three years of the survey.  Female leaders are shown in red and male leaders in purple and error bars show the propagated Poisson error.
       \label{fig:gender_daw}
       }
\end{figure}

We find no significant differences between the fractions of male and female leaders across all three years for four of the five paths; the exception is the \texttt{defined around work I was already doing} category.  
\autoref{fig:gender_daw} shows the gender-normalized results, meaning the positive responses for female (male) over the number of female (male) leaders, for the \texttt{Defined Around Work} path to leadership.  
We find that females are less likely than males to have leadership roles defined around their existing work.  
Furthermore the gap between men and women grows larger over time, with the fraction of men continuing to hold roles defined around their work at the same rates, while decreasing for women.        

\section{SDSS as an inclusive environment} \label{sec:surveyclimate}

In the 2015 demographic survey we added a question about the general climate of the collaboration. Specifically we stated: \texttt{The SDSS fosters an inclusive climate} and respondents could respond: \texttt{strongly disagree}, \texttt{disagree}, \texttt{neutral}, \texttt{agree}, \texttt{strongly agree}, \texttt{unsure}, or \texttt{prefer not to answer}. In 2015, 15\% \texttt{strongly agreed} and 47\% \texttt{agreed} and in 2016 17\% \texttt{strongly agreed} and 58\% \texttt{agreed}. Overall, more than 60\% of respondents \texttt{agreed} or \texttt{strongly agreed} in both years, which indicates that the SDSS-IV collaboration is perceived to be inclusive.  

\begin{figure}[h]
        \includegraphics[width=1.0\textwidth]{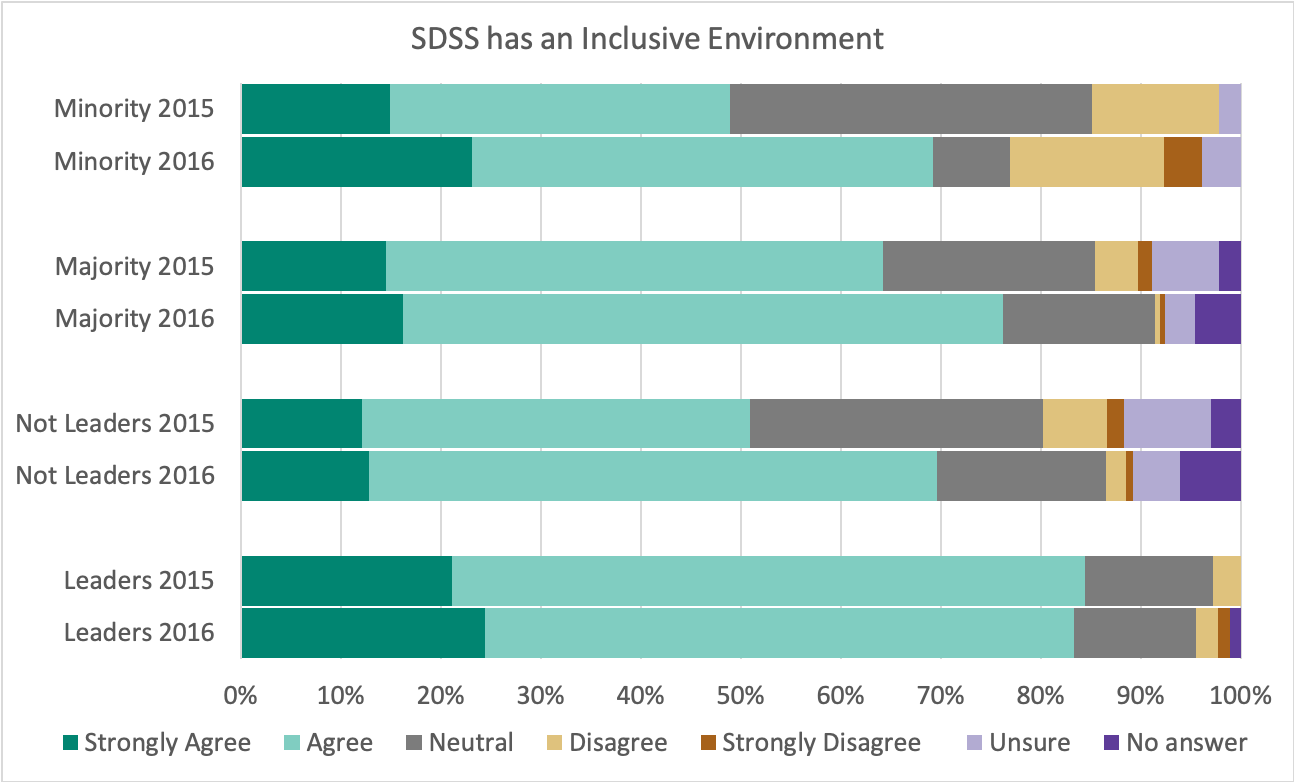}

    \caption{Responses to \texttt{The SDSS fosters an inclusive climate} compared amongst some demographic groups from 2015 and 2016 surveys. Respondents could select \texttt{strongly agree} (dark green), \texttt{agree} (light green), \texttt{neutral} (grey), \texttt{disagree} (light brown), \texttt{strongly disagree} (dark brown), \texttt{unsure} (light purple), and \texttt{prefer not to answer} (dark purple), shown here from left to right, respectively. 
    The top two bar graphs compare minority (top) and majority (second from top) groups. 
    Not Leaders (second from bottom) and leaders (bottom) are compared on the bottom two bar graphs. Counts are not provided due to small response rates in some of the bins. \label{fig:min_lead_incl}}
\end{figure}

When we examine the answers within different demographic groups, the fraction of people that (dis)agree shifts. 
It is important to better understand these shifts and which demographic groups find SDSS to be less inclusive.  The largest differences among various demographic groups were found when comparing people from self-reported minority and majority race or ethnic groups (see \autoref{fig:demographics_prt1}b for the definition of minority/majority groups in the international SDSS collaboration), and self-reported leaders and not leaders (see \autoref{sec:leadership_def} for definition of leader). These data are shown in \autoref{fig:min_lead_incl} for years 2015 and 2016. 
We will focus the discussion on these since they had the strongest contrast, noting however that there were other differences seen and many demographics overlap.

In general, people from minority race or ethnic groups report finding SDSS to be less inclusive than those from the majority group.  In particular, in 2015 less than 50\% of minorities agreed or strongly agreed compared to 64\% for majorities.  Also 36\% felt neutral about SDSS fostering an inclusive environment.  A higher percentage of minorities in both years disagreed in contrast with the majority (13 and 19\% of minorities compared to 6 and 1\% of majorities for 2015 and 2016, respectively).  Even though from 2015 to 2016 more minorities thought SDSS fostered an inclusive environment, there was also a higher fraction that strongly disagreed.  This contrast between self-reported minorities and majorities illustrates that more effort should be made to better foster an inclusive environment for minorities within SDSS.  

The other large difference between two demographic groups was seen between self-reported \textit{leaders} and \textit{not leaders}.  The lower bar charts in \autoref{fig:min_lead_incl} show that self-reported \textit{leaders} are more likely to agree with the statement that the SDSS-IV fosters an inclusive climate.  Over 20\% (21 and 24\% in 2015 and 2016) of \textit{leaders} strongly agree compared to 12 and 13\% (in 2015 and 2016) of \textit{not leaders} and over 80\% of \textit{leaders} are in agreement versus 51 and 70\% in 2015 and 2016. Also \textit{leaders} were more likely to have an opinion on climate compared to \textit{not leaders} (i.e., there are no \texttt{unsure} and less than 1\% were \texttt{no answer} answers among leadership). These differences between \textit{leaders} and \textit{not leaders} illustrate that it is important to survey everyone within SDSS and not just leadership to avoid biased results. 
    
\section{Recommendations and Future Work} \label{sec:summary}

For each of the previous (sub)sections, we provide a summary, conclusion and recommendations.  The recommendations are aimed towards the astronomy community as a whole, in particular any large collaborations like SDSS. We will quote the minimum and maximum of any numerical results from the three years of the survey in this section.  We end with a discussion about future work.

\subsection{SDSS-IV Demographics}

\textit{Summary:} In  \autoref{sec:overalldemo} we report the overall demographics of members of SDSS-IV across the three years of the survey, 2014-2016. 

\textit{Conclusion:} SDSS-IV is an international collaboration: we find that in all three years, about half of respondents are based in North America, a quarter in Europe, and the remainder in Asia and Central and South America. Overall we find 11-14\% of members are racial or ethnic minorities where they live, 26-36\% are women and up to 2\% report non-binary genders. These proportions are compared to those from AAS, RAS and (where available) IAU, and we find they are mostly consistent with the demographics of members of these groups reported at similar times. 32 to 43\% of respondents were within five years of receiving their terminal degree. Similarly about half are in a junior career role, with senior faculty decreasing from 27 to 20\%.

\textit{Recommendations}: We find that the demographics of the SDSS-IV membership is similar to other groups of astronomers, and no significant changes have been seen over the three years of data we report here. SDSS-IV is a biased sample of astronomers, since resources (money or time) were required to gain access to the collaboration. Collaborations based on buy-in should consider how that impacts their demographics. 

Astronomy as a whole, and the SDSS-IV collaboration remains dominated by white heteronormative men, although not to the extent it was decades ago. We recommend that large astronomical collaborations, in common with other organizations, commit resources to track the demographics (including gender, race and other metrics) of their membership over time. 

\subsection{Gender Balance Worsens with Academic Age}

\textit{Summary:} In \autoref{sec:gender}, the representation of women and men\footnote{The SDSS-IV demographic survey included options to report non-binary genders; however the number of collaboration members choosing such options is small, so for the sake of anonymity we are unable to include them in further analysis.} in the survey were compared as a function of academic age tracked either by years since terminal degree, or career role.   

\textit{Conclusion:} We find larger fractional representation of women at ``younger'' academic ages, with $35-47$\% of student respondents being women but just $20-25$\% among the most experienced astronomers. This same trend is seen with career role. This phenomenon, often called the ``leaky pipeline", a term which describes a decreasing representation of women with career stage or academic age, is also seen among the AAS Workforce. We note that there has been an increase in the number of women being awarded astronomy degrees and it may take more time for this increase to progress through the pipeline. We conclude that there remains a signal of the ``leaky pipeline'' among astronomers in the SDSS-IV collaboration, i.e. that women are more likely than men to leave over time. 

\textit{Recommendations}: There is evidence that women experience more barriers to success in research careers than men, on average. When identifying excellence in science and/or scientists we all should be aware of the impact of ``opportunity bias", and should seek {\it excellent potential} rather than the biased metrics of success. There also should be consideration to providing additional support at common exit points (e.g. childcare at conferences, allowance for the extra burden of other caring responsibilities which often falls on women) to make remaining active in research more possible. 

\subsection{Education/Socio-Economic indicators}

\textit{Summary:} Measures of the educational background of SDSS-IV members as a way to capture information on socio-economic background were studied in \autoref{sec:edubkgd}. These data were available in 2015 and 2016 surveys only. We report the fraction of SDSS-IV members with no parents with a college degree, at least one parent with a college degree, and at least one parent with education beyond a college degree split by masters, professional degrees, and Ph.Ds. We have sufficient survey responses to additionally break this question down by gender in \autoref{sec:edugender}. In \autoref{sec:fgli} we regroup data in a way to identify first-generation college students (FGCS; meaning members with no parents with a college degree), which gives sufficient sample sizes to break the question by both race/ethnicity and gender.  

\textit{Conclusion:} We find that SDSS-IV members overall are nearly twice as likely to have a parent with any college degree than the general population and {\bf ten times} more likely to have a parent with a PhD than the general population (in all countries where SDSS-IV members live/work). While no comparison data is available for the astronomical workforce, we do find data showing a similar trend for academic faculty in general (that they are much more likely to have parents with advanced degrees). Breaking this down by gender in the SDSS-IV collaboration, we find that these trends are amplified among women in the collaboration (a 1$\sigma$ increase). This result was previously reported in \citet{lucatello_2017} who suggested it was evience that women need higher educational background/socio-economic status to persist in astronomical/science careers than men.   

Looking at the fraction of FGCS in SDSS-IV, we find that overall it is 31\%. This fraction increases (with low statistical significance due to the small numbers) to $40-50$\% among SDSS-IV members who identify as being part of a minority racial group, and decreases (more than $1\sigma$) to $15-25$\% for women. Women FGCS have an intersectional identity which appears to make it even harder to persist in STEM than women in general. And scientists from minority racial/ethnic groups may be more likely to be FGCS than majority scientists, suggesting that making sure FGCS scientists have the support they need may help increase the diversity of the astronomical workforce. 

\textit{Recommendations}: Academics as a whole (including astronomers) generally come from highly educated backgrounds. However a substantial fraction of astronomers are FGCS, who may need additional support navigating the ``hidden-curriculum" of academic careers {\bf at all stages}. Since there is evidence that people who identify as being part of a minority racial group are more likely to be FGCS than the majority group, providing this support to promote retention of FGCS in STEM careers, may also help to increase racial diversity. Additionally we find evidence that women from less educated family backgrounds may find it harder to persist in STEM/astronomy careers than men from similar backgrounds (i.e., the gender balance among FGCS is worse than overall), so FGCS women may need additional support to be successful.

\subsection{Gender imbalance in leadership, recognition of leadership and routes to leadership}

\textit{Summary:} In \autoref{sec:leadership} we report the demographics of people in leadership roles in SDSS-IV (in \autoref{sec:leadership_def} we define what that meant in the collaboration). We looked at the data on these roles as a function of gender, career-stage/age, previous SDSS membership, and social status, however due to sample size, the only significant trends related to gender differences, so these are the only trends we discuss. In addition to data from the survey, we also looked at the gender breakdown of people listed on organizational charts of high-level leadership roles in SDSS-IV, and also the different routes people take into leadership roles in the collaboration. 

\textit{Conclusion:} Overall we find that about 30\% of respondents report some kind of leadership role in SDSS-IV. We find that men who are part of the SDSS-IV collaboration are more likely than women to self-report they hold leadership roles, have recognized leadership roles (i.e. named on an organizational chart), and be more likely to be in leadership roles associated with funding. When we normalize over gender, the difference between male and female respondents is significant for both recognized and funded leaders, with women having a smaller fraction compared to men. However after removing the gender bias within leadership (i.e., normalize by male/female leaders), the difference is less significant. Looking at named ``key personnel", we found the gender imbalance, compared to the overall gender breakdown of survey respondents was particularly large (23\% of them are women), unless they are compared to only the most senior members of the collaboration (e.g. 24\% of people 16+ years from final degree were women in 2016 data).  

SDSS-IV has worked hard on encouraging open calls for leadership roles, with a formal policy for advertising. Despite this we found that the most common ``path to leadership" from our survey was more informal, via people being directly asked to fill a roll, or having a roll designed around existing contributions. These routes appear to propagate gender biased expectations; we show evidence that men in the collaboration were more likely to find a leadership roll designed around their existing work than women. In 2016 this difference was significant, even after normalizing by male/female leaders. Roughly half of male leaders become a leader defined around work they were already doing compared to less than a third of female leaders. 

\textit{Recommendations}: A policy of requiring open calls for leadership is a good first start, but is clearly not enough to lower the additional societal barriers women may face being in a position to be ready for formal or informal leadership roles in a science collaboration. Attention needs to be paid to the kind of work junior women in a collaboration are asked to do, to consider if those contributions may be considered less valuable as leadership skills than others (e.g. contributions to technical pipelines, or instrument building, compared to education or communication roles). Formal set-up of mentoring for both new leaders, and potential future leaders may also help.   

\subsection{Perception of Inclusion}

\textit{Summary:} In \autoref{sec:surveyclimate} we consider answers to a question about the level to which respondents felt SDSS-IV fostered an inclusive climate in the collaboration. This question was introduced in the 2015 survey, and repeated in 2016. 

\textit{Conclusion:} Overall we conclude SDSS-IV is perceived as an inclusive collaboration by it's membership -  more than 60\% of all respondents in both 2015 and 2016 agreed (or strongly agreed) with this statement and there was an increase in the fraction agreeing between the two surveys. However, notable differences were seen between answers from respondents who self-reported being part of a racial or ethnic minority group where they work. Members of a minority group are less likely to agree with the statement on inclusiveness than majority group members; although in both groups the fraction increased between the two surveys. We also find that people in leadership roles were notably more likely to agree with the statement than those not in leadership roles. 

\textit{Recommendations}: While overall SDSS-IV is perceived as inclusive, more work can be done to foster an inclusive environment for minority racial or ethnic group members. It is also crucial to survey everyone in a collaboration, not just leadership, on the success of inclusive practices. 

\subsection{Future Work} \label{sec:future}

In addition to the three demographic surveys used here, COINS conducted a Demographic Survey of SDSS-IV in both 2018 and 2021. 
The full demographic dataset from SDSS-IV thus spans seven years with five individual surveys. 
This paper has primarily focused on demographic axes first probed by \citet{lundgren_2015} and revisited in commentary by \citet{lucatello_2017}; as can be realized via a skim of the full 2016 survey in \autoref{app:survey}\footnote{The questions for all five surveys can be found on the COINS Github at \url{https://github.com/sdss/coins/tree/main/documents/md}.}, there is much more information in the surveys than has been analysed here. 
Thus, future studies may both cover the full seven years of data and explore additional demographic axes covered by the survey.

The 2021 survey included questions relating to COVID19 impacts.
Additionally, in collaboration with the Organizing Committees of the annual SDSS Collaboration meeting, COINS has also surveyed SDSS-IV members regarding their impressions of collaborative practice during conferences. 
Thus, future work may explore these aspects of large-collaboration culture and inclusive practice.

As in prior transitions between phases of SDSS, SDSS-IV and SDSS-V operated simultaneously for a period of time. 
Thus, the 2021 Demographic survey included members of SDSS-V. 
Many of the policies described in \autoref{app:current} and referenced throughout this work were implemented after some decision-making at the project level in SDSS-IV. 
In contrast, these polices were a component of the formative processes for SDSS-V. 
Future work will be able to examine differences in representation after such policies are adopted, which may be impactful for the future of astronomy with larger collaborative efforts serving an international community.

\acknowledgements
\emph{Acknowledgments:} 

The Committee on INclusiveness in SDSS (COINS) warmly thank the many collaboration members who completed the Demographic Surveys. We also thank all the current and past members of COINS, the Committee on the Participation of Women in SDSS (CPWS), the Committee on the Participation of Minorities in SDSS (CPMS), and the SDSS Management Council for all their incredible work and efforts to enable these surveys and support improving the SDSS and scientific culture. Amy Jones was partially funded by STScI DDRF D0001.82526 project entitled ``Inclusion in International Projects: Exploiting a Decade of Demographic Surveys in SDSS-IV to Inform Science Operations''.

Funding for the Sloan Digital Sky 
Survey IV has been provided by the 
Alfred P. Sloan Foundation, the U.S. 
Department of Energy Office of 
Science, and the Participating 
Institutions. 

SDSS-IV acknowledges support and 
resources from the Center for High 
Performance Computing  at the 
University of Utah. The SDSS 
website is www.sdss4.org.

SDSS-IV is managed by the 
Astrophysical Research Consortium 
for the Participating Institutions 
of the SDSS Collaboration including 
the Brazilian Participation Group, 
the Carnegie Institution for Science, 
Carnegie Mellon University, Center for 
Astrophysics | Harvard \& 
Smithsonian, the Chilean Participation 
Group, the French Participation Group, 
Instituto de Astrof\'isica de 
Canarias, The Johns Hopkins 
University, Kavli Institute for the 
Physics and Mathematics of the 
Universe (IPMU) / University of 
Tokyo, the Korean Participation Group, 
Lawrence Berkeley National Laboratory, 
Leibniz Institut f\"ur Astrophysik 
Potsdam (AIP),  Max-Planck-Institut 
f\"ur Astronomie (MPIA Heidelberg), 
Max-Planck-Institut f\"ur 
Astrophysik (MPA Garching), 
Max-Planck-Institut f\"ur 
Extraterrestrische Physik (MPE), 
National Astronomical Observatories of 
China, New Mexico State University, 
New York University, University of 
Notre Dame, Observat\'ario 
Nacional / MCTI, The Ohio State 
University, Pennsylvania State 
University, Shanghai 
Astronomical Observatory, United 
Kingdom Participation Group, 
Universidad Nacional Aut\'onoma 
de M\'exico, University of Arizona, 
University of Colorado Boulder, 
University of Oxford, University of 
Portsmouth, University of Utah, 
University of Virginia, University 
of Washington, University of 
Wisconsin, Vanderbilt University, 
and Yale University.


\bibliographystyle{aasjournal}
\bibliography{00_bib}

\appendix

\section{A History of Recent Inclusion Efforts within SDSS}\label{app:history} 

A central objective of administering an annual/biannual demographic survey is to monitor the impact of equity and inclusion efforts within SDSS. 
Thus, the following brief history of these efforts within SDSS is relevant to interpret the results of the survey.

The original {\it Committee on the Participation of Women in SDSS} (CPWS) operated from October 2012 to July 2013 within SDSS-III \citep{eisenstein_2011} and was initiated in response to a recommendation from the Sloan Foundation to evaluate the demographics of the collaboration, with a particular focus on the participation of women in leadership positions.
The specific charges to the CPWS were:
(i) examining how collaboration leadership is established, 
(ii) evaluating the diversity and general climate within the collaboration, with a particular emphasis on gender balance, 
(iii) fielding climate-related concerns from people within the SDSS, 
and (iv) making recommendations to the SDSS management on how to improve the overall equity and climate within the collaboration.

The CPWS became a standing committee in SDSS-IV in December 2013, with annual reports provided in \citet{CPWS_2015AAS,CPWS_2016AAS}. 
The CPWS administered the first collaboration-wide, voluntary demographic survey in April 2014, which corresponded to the ``initial'' demographic state of SDSS-IV.
The original demographic survey was designed to take a ``baseline'' demographic snapshot against which changes could be compared and evaluated in the future.

In 2014, SDSS underwent two external reviews.
First, the Statistical Research Center of the American Institute of Physics, in conjunction with SDSS, adapted a questionnaire to evaluate climate that better suited the collaboration environment rather than the university or research lab environment (e.g., the environment governing other questionnaires of this type). 
Responses were collected through June 2014. 
At the 2014 Collaboration Meeting in Park City, Utah, the Committee on the Status of Women in Physics of the American Physical Society (APS) conducted a ``site visit'' that included interviews as well as observations of participation during the meeting. 
Reports from both activities were issued thereafter.
The survey suggested that most respondents felt like the collaboration worked well, but some significant differences were noted between men and women, including men sensing fewer barriers to leadership and women feeling more pressure to attend all meetings.
The report on the site visit included actionable interventions based on best practices from other large collaborations in particle physics, which included a code of conduct, ombudspeople, and proactive efforts to welcome new members, among others.
As will be described in \autoref{app:current}, these reports were very influential.

In the Spring of 2015, the CPWS ran a second demographic survey, which included all the questions from the previous survey for a direct comparison of responses.
New questions were added to expand the scope of the demographic portrait with the collection of information related to other components of diversity, including race, sexual orientation, marital status, and family background; the motivation for this is well summarized by \citet{prescod-weinstein_2017} and insights from these additional demographics axes were discussed in \cite{lucatello_2017}.
Moreover, members were asked to provide feedback about their perception of the SDSS-IV climate and of the opportunities it affords to themselves and others.

Parallel with CPWS, in 2012 the {\it Committee on the Participation of Minorities in SDSS} (CPMS) was formed with the charge of recommending and implementing strategies to increase the participation of underrepresented minorities in the SDSS collaboration. 
This led to the creation of the {\it Faculty and Student Teams}\footnote{ \url{https://www.sdss.org/education/faculty-and-student-team-fast-initiative/}} (FaST) initiative that pairs professors and students at minority serving institutions directly with SDSS scientists and affords full access to SDSS data products.
The FaST participants become ``full'' collaboration members.

In 2016, the CPWS and the CPMS were merged to form the {\it Committee for INclusion in SDSS} \citep[COINS;][]{schmidt_2017_coins} which continues to support the keystone initiatives from the original committees. 
The formal goal of COINS is to assess the climate and demographics of SDSS, to recommend new policies or practices concerning increasing inclusiveness, and to assist in the implementation of these new activities where necessary. 
Thus, COINS continues to administer a routine demographic survey and supports the FaST program.
Additionally, the committee works to improve interactions between members in SDSS by making active recommendations for the day-to-day actions of the collaboration.

COINS operates within the Collaboration Meetings to engage new members and student participants by planning events during the meeting that ensure these individuals interact broadly at the meeting. 
COINS hosts regular ``town hall'' style meetings where SDSS members can discuss concerns; these are well attended by SDSS-IV leadership, which provides direct access to these individuals. 
COINS has produced a set of recommendations for 
(i) inclusive teleconferencing, which includes possible solutions to timezone exclusion and encouraging active participation from new attendees,
(ii) accessibility at meetings, 
(iii) accessible and inclusive session chairing, 
(iv) and a ``Best Practices'' document that accompanies the SDSS Code of Conduct.\footnote{The Code of Conduct is available at the following URL: \url{https://www.sdss.org/collaboration/the-sloan-digital-sky-survey-code-of-conduct}}

Similar to how SDSS makes its data products public and its pipelines open access, the documentation produced by COINS are also now included on the SDSS website and are updated with each data release\footnote{Available: \url{https://www.sdss.org/collaboration/coins/}}.
Further, COINS has created a GitHub Repository, alongside SDSS software repositories, that contains COINS-generated content, including policy documents, presentations, and materials for activities.\footnote{\url{https://github.com/sdss/coins/}}

\section{SDSS-IV Collaboration Policies} \label{app:current}

{\it Management framework.}
The principles of operation are publicly available\footnote{ \url{https://www.sdss.org/wp-content/uploads/2014/11/principles.sdss4_.v4.pdf}} and define the operating policies of the project. 
The management of SDSS occurs via a set of standing committees: the Advisory Council, the Management Committee, and the Collaboration Council act on a survey-wide level with representation drawn from and selected by member institutions.
Policies for recognition of effort are described as are policies regarding publications, including internal review periods and authorship policies.

{\it A set of Ombudspeople.}
The ombudspeople are available anonymously and independently of management to advise collaboration members on any concerns/questions they may have. 
This role was established after the 2014 site visit by the AIP. 
At least two people serve in this capacity at all times and this increases the effectiveness of this role.
The COINS Github has a space dedicated to the Ombuds Office.\footnote{\url{https://github.com/sdss/coins/tree/main/documents/md\#ombuds-office}}

{\it A standing committee operates to evaluate inclusion in SDSS.}
As described in the previous subsection, the initial CPWS and CPMS were merged in 2016 to form COINS \citep{schmidt_2017_coins}.
Thus, a committee has been in place since the outset of the SDSS-IV project. 
The committee holds regular town halls at collaboration meetings and regularly cycles membership (though there are no committee term limits).

{\it SDSS management positions are filled akin to formal hiring.} 
Leadership positions within SDSS follow the same cycle as formal hiring, with an open call, a formal announcement, and active recruiting. 
See \autoref{sec:fillingroles} for the full policy document

{\it Leadership roles and terms are clearly defined.} 
The responsibilities of leadership positions are described on the internal wiki.
Many leadership positions are held in a ``rotating'' status, which provides more access to leadership roles. 
See \autoref{sec:fillingroles} for the full policy document.

{\it Integration of new members as a conscious effort.}
Throughout the internal SDSS pages, efforts are made to ``welcome'' new members into active research, including definitions of jargon, quick-start guides, etc. 
Efforts are made at collaboration meetings to make leadership visible and accessible, as well as specific activities to help new members meet other members.
Lastly, data analysis and pipeline teams host ``Q\&A'' efforts with respect to data releases and documentation efforts.

{\it Code of Conduct.}
In 2018, SDSS-IV adopted a {\it Code of Conduct} for participation in the project; this document is maintained publicly.\footnote{\url{ https://www.sdss.org/collaboration/the-sloan-digital-sky-survey-code-of-conduct/}}
The {\it Code of Conduct} was developed from 2015 to 2016 via a committee. 
Several town-halls and open discussions were held on its specific wording and implications from 2016 to 2017.
Final adoption occurred in early 2018. 
All participants in SDSS-IV agreed to follow the {\it Code of Conduct}. A {\it Best Practices} document was developed by COINS to place the code into the context of typical interactions for collaboration members.

\subsection{SDSS-IV Practices Document: Regarding Filling Leadership Roles in the Project} \label{sec:fillingroles}

Below is a practices document that was produced to outline how leadership roles should be filled in SDSS-IV that was posted to the SDSS-IV wiki alongside other policy documents. 
\begin{quote}
\noindent {\bf SDSS-IV Practices Document} \\
{\bf Regarding Filling Leadership Roles in the Project} \\
{\bf Michael Blanton} \\
{\bf September 11, 2013} \\

This document describes how we approach filling leadership and technical roles in the survey. This includes Management Committee roles, as well as Science Collaboration roles such as Science Team Chairs and Working Group leads. The primary purpose of these practices is to ensure an open process that results in project management which is representative of the collaboration, in particular in terms of gender balance.

{\bf Tracking and approval:} The leadership and technical roles will be tracked on the project wiki, along with brief descriptions of their roles and responsibilities. Such roles will be recommended by the survey teams and approved by the Director. For each such position or committee filled, SDSS-IV leadership will track the information on the number of candidates as well as the gender information regarding the candidates and the filled positions. This information allows the project to track its progress.

{\bf Duration of appointment:} Leadership roles will have specified durations of appointment. While some positions we will expect to continue to be filled by the same person throughout the project, most positions will have shorter terms. The expectation for such positions is that they will be turned over to a new person, though if appropriate the same person can continue for multiple terms.

{\bf Advertisement:} All openings for leadership positions should be announced to the collaboration, with a description of the responsibilities and expectations for the position. Some information about the position's compensation should be given, though it should be clear that compensation will depend on experience, level of commitment, and other contingencies. Other benefits of the position are useful to mention, in particular if lead authorship on a technical paper, Architect status, or other intangible benefits are likely to result. Members of the collaboration should be encouraged to express interest in the position to a designated contact who can answer any other questions about the position.

{\bf Targeting likely candidates:} It is acceptable for the groups making recommendations to identify likely candidates and recruit them to apply to these positions. We particularly encourage these groups to focus on identifying female candidates to apply.

{\bf Consideration of applications:} Prior to advertisement, the group charged with making a recommendation for appointment will discuss the qualifications that they are looking for in successful candidates. The group making the recommendation is required to submit their candidate list to the SDSS-IV MC Executive Committee (Director, Program Manager, Project Scientist and Project Spokesperson). The MC
Executive Committee will act to ensure that the group has made sufficient effort to develop a candidate list with a balanced gender distribution.

Exception to this practices document can be made at the discretion of the Director.
\end{quote}

\section{2016 Demographics Survey} \label{app:survey} 
The questions from the 2016 survey are presented in the panels of \autoref{fig:2016_survey_questions} and also on the COINS Github\footnote{\url{https://github.com/sdss/coins/blob/main/documents/md/demographics_survey_2016.md}}.
The questions from all surveys issued in SDSS-IV can also be found on the COINS Github. 

\begin{figure*} \centering 
 \begin{center}
  \includegraphics[angle=0,width=3.2in]{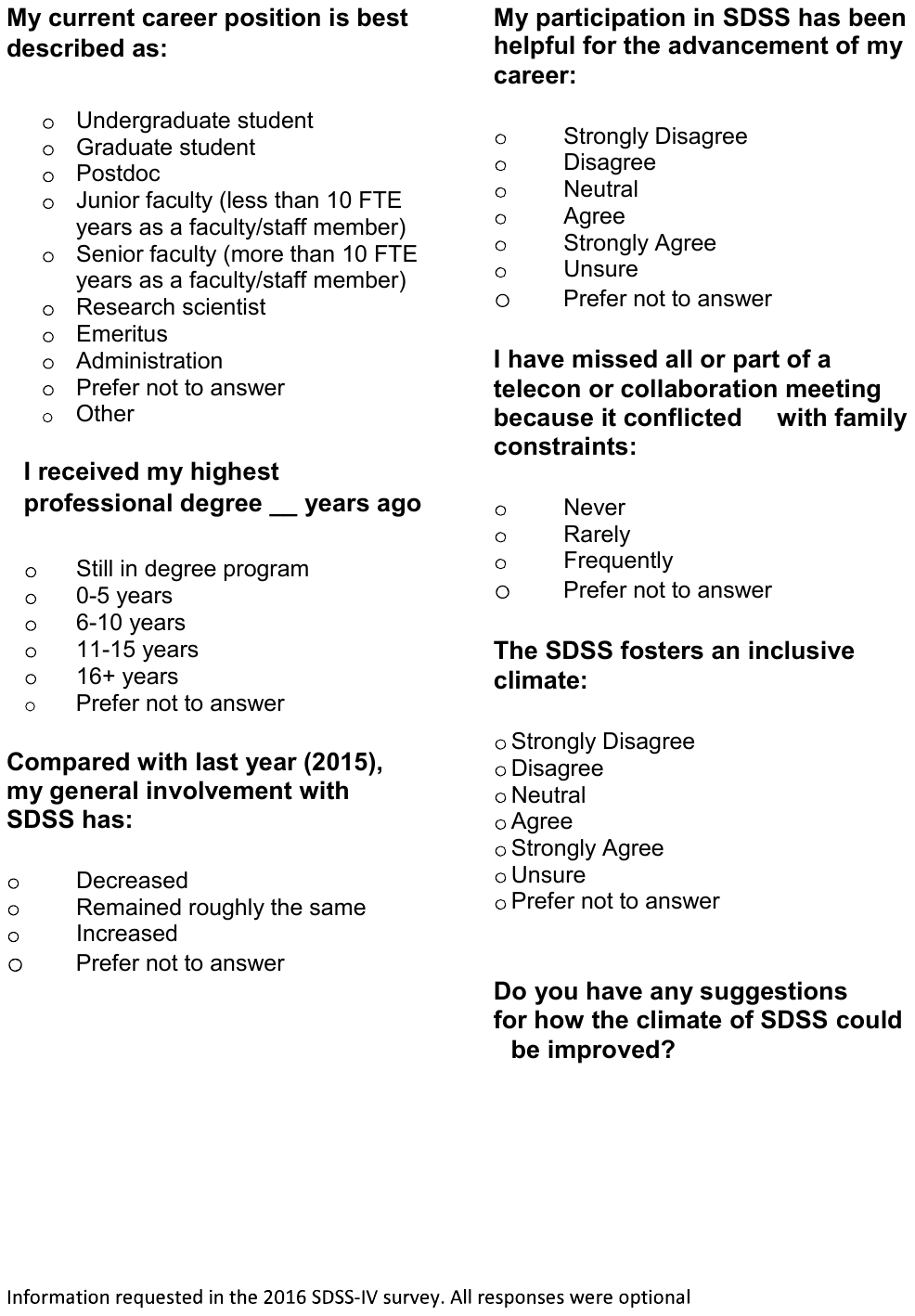}
  \includegraphics[angle=0,width=3.2in]{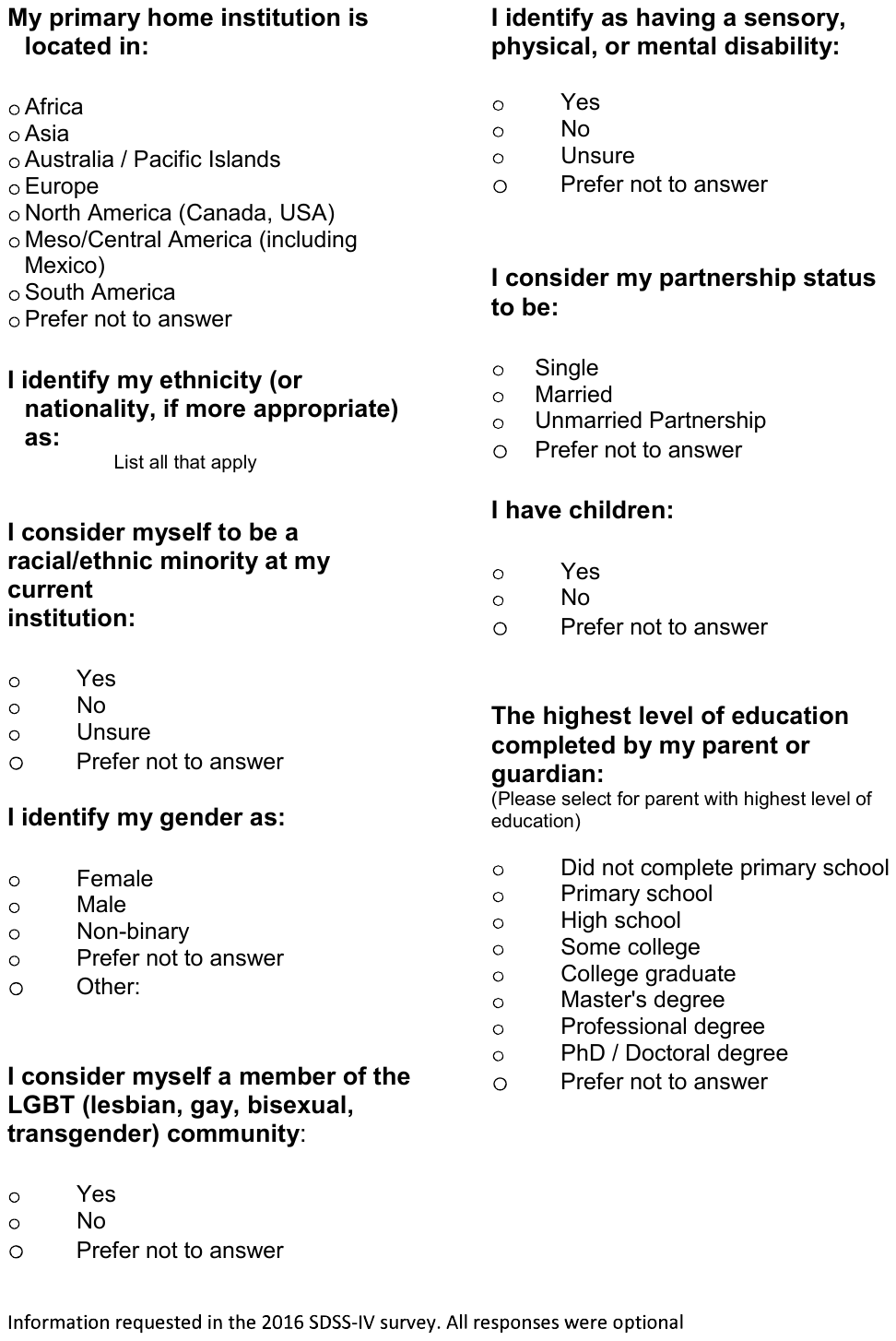} \\
  \includegraphics[angle=0,width=3.2in]{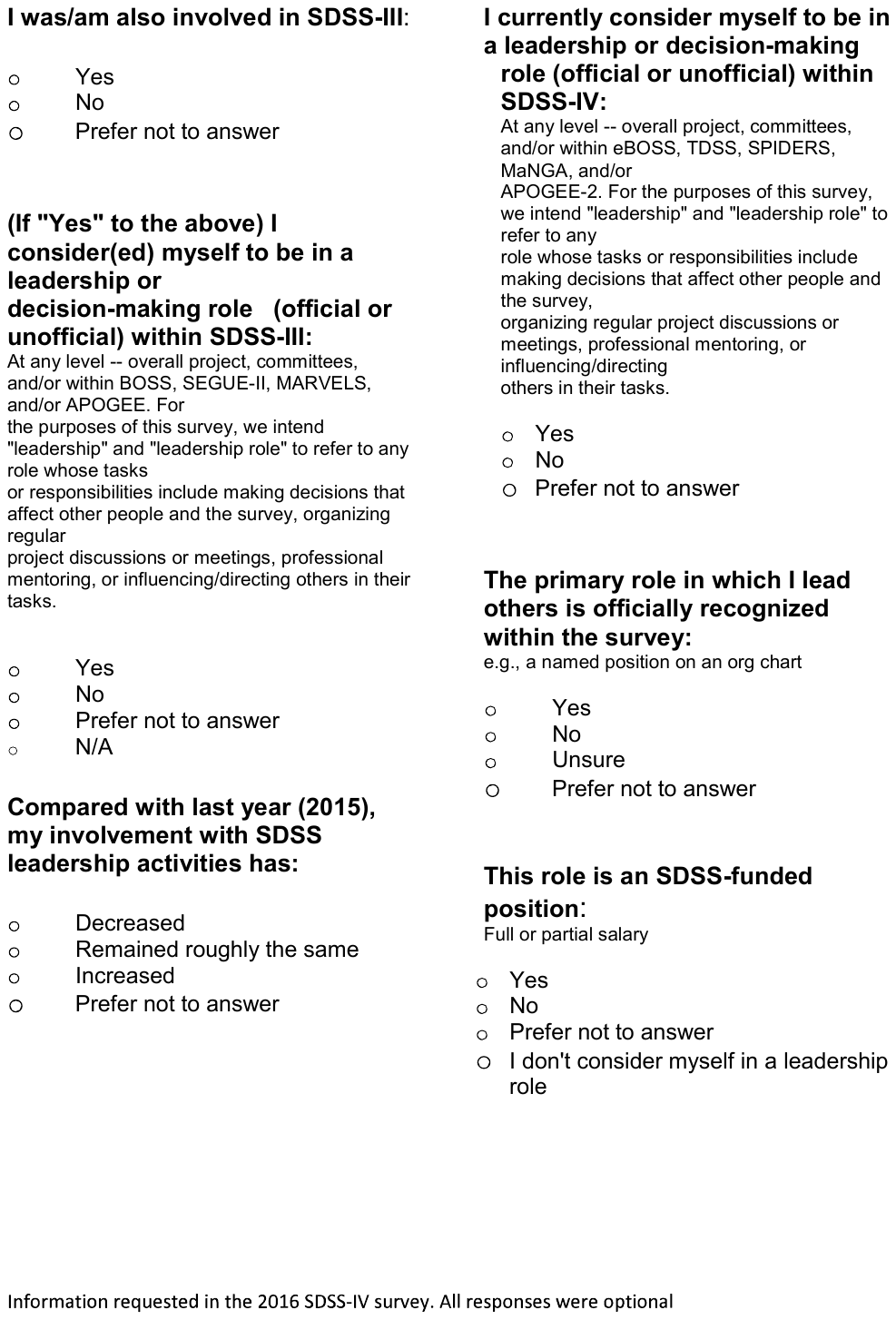}
  \includegraphics[angle=0,width=3.2in]{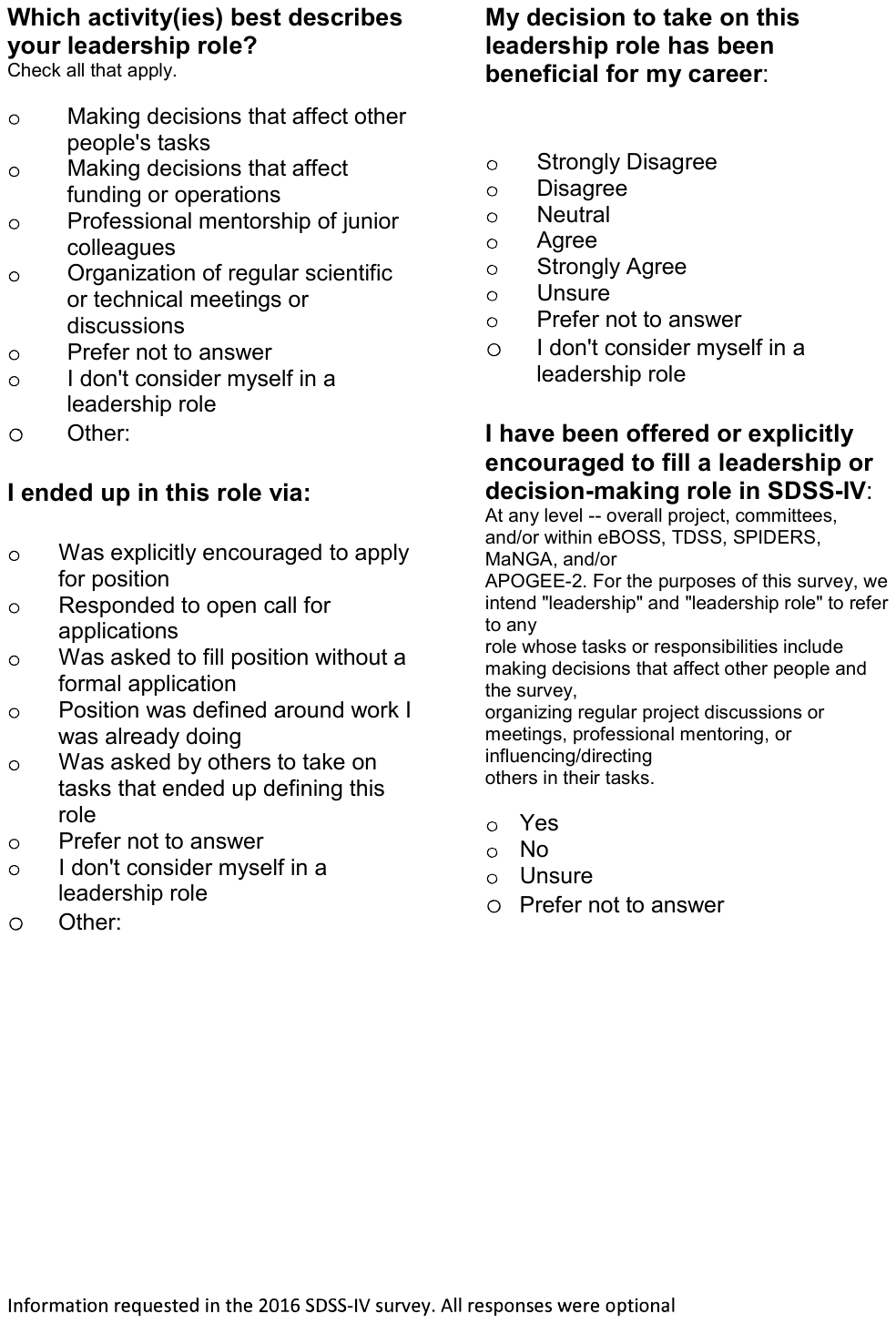}
 \caption{Information requested in the 2016 SDSS-IV survey. All responses were optional. \label{fig:2016_survey_questions}}
\end{center}
\end{figure*}
\setcounter{figure}{8}
\begin{figure*} \centering 
 \begin{center}
  \includegraphics[angle=0,width=3.2in]{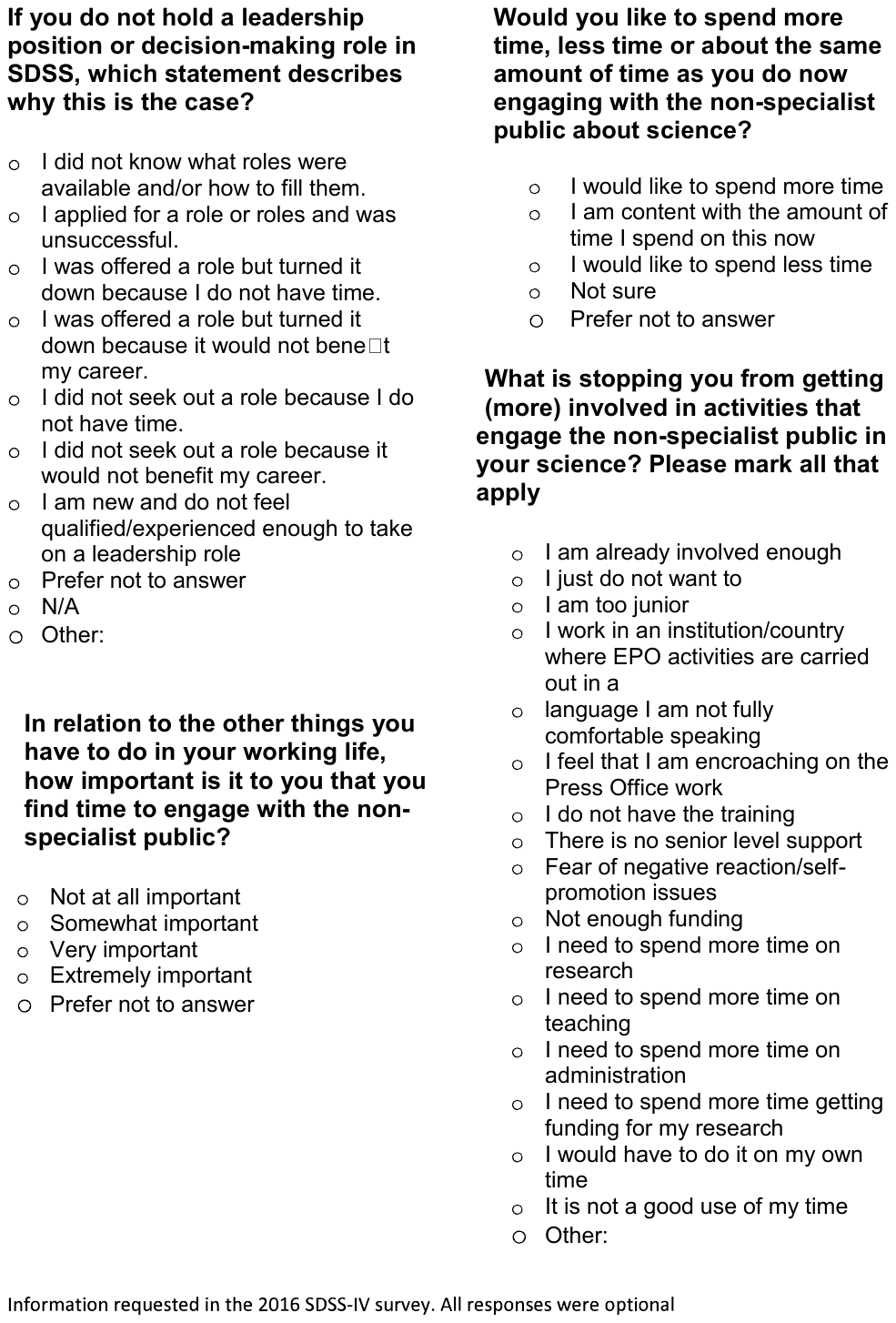}
  \includegraphics[angle=0,width=3.2in]{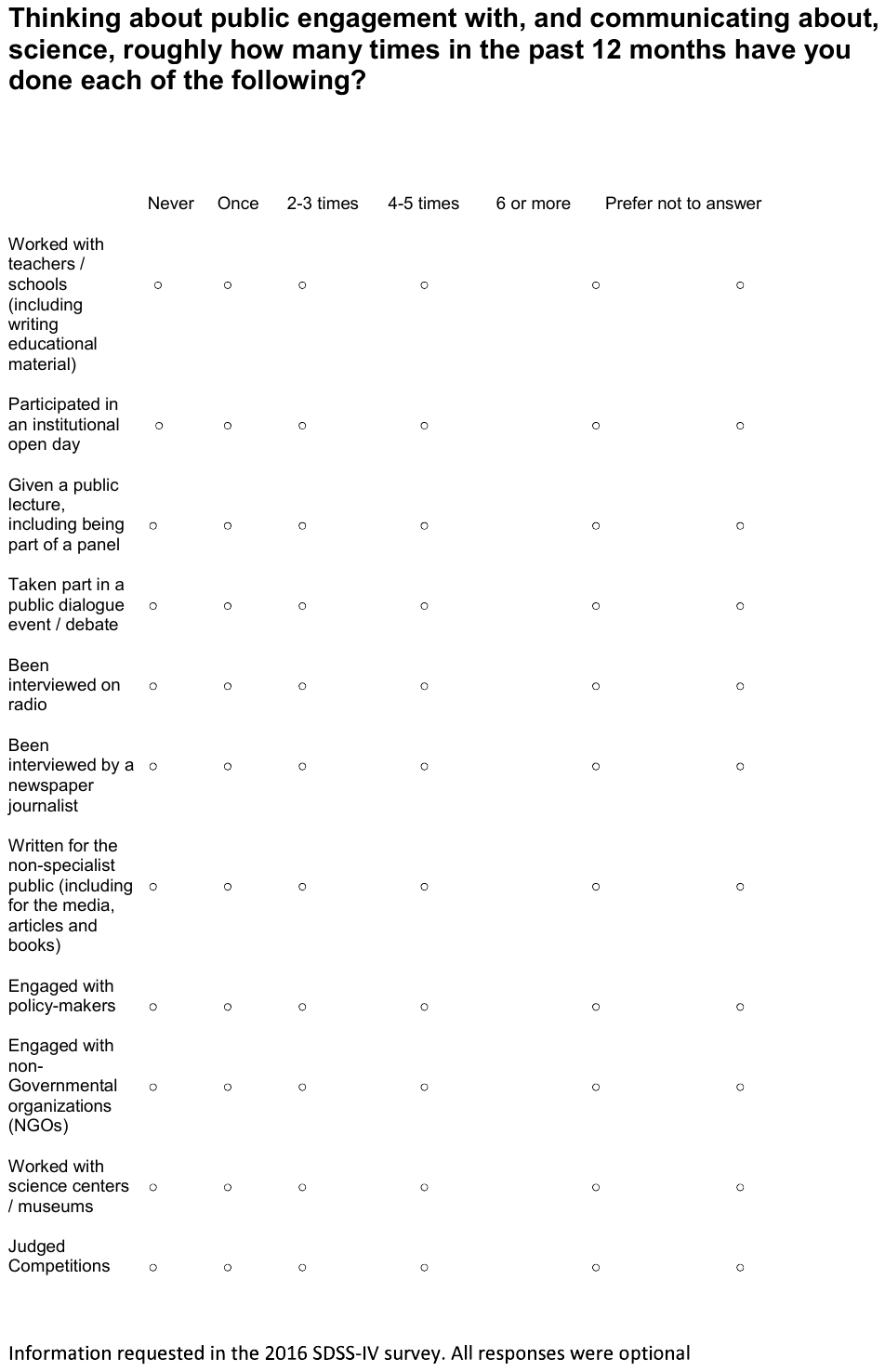}
 \caption{ -- cont. }
 \end{center}
\end{figure*}

\begin{table}
    \centering
    \caption{Results from the three survey years for respondents  who  self-recognized  as  having  a  leadership  role  in  the  SDSS-IV,  first broken down by academic age given as the number of years since they received their highest professional degree for women (F) and men (M) and then by career stage.}
    \label{tab:lead_gend_age}
    \begin{tabular}{lrrrrrr}
    \hline
    & \multicolumn{2}{c}{LEADERS} \\
     & \multicolumn{2}{c}{2014} & \multicolumn{2}{c}{2015} & \multicolumn{2}{c}{2016} \\
     & F &	 M	& F	&  M & F	&  M \\
    \hline
    \hline
    {\bf Academic Age}:\\
    $<5$ yrs     & $<5$ & 15 & 6 & 24 & 8 & 17 \\
    $6-10$ yrs     & 7 & 15 & 6 & 19 & 6 & 14 \\
    $11-15$ yrs    & $<5$ & 13 & 5 & 18 & $<5$ & 10 \\
    $16+$ yrs      & 6 & 25 & 7 & 24 & 8 & 21\vspace{.5em}\\

    {\bf Career Stage}:\\
    Grad+Postdoc & $<5$ &  9 & 6 & 22 & 9 & 13\\
    Junior Faculty & 5 & 16 & 6 & 15 & 8 & 16\\
    Senior Faculty & 6 & 26 & 9 & 24 & 7 & 17\\
    Research Scientist & 5 & 14 & $<5$ & 18 & $<5$ & 10\\
    Administration/Other & $<5$ & $<5$ & $<5$ & 6 & $<5$ & 6 \\
    \hline
    \end{tabular}
\end{table}

\begin{table}
    \centering
    \caption{Gender breakdown of self-reported leaders with other demographic factors.}
    \label{tab:lead_other}
    \begin{tabular}{lrrrrrr}
    \hline
      & \multicolumn{2}{c}{2014} & \multicolumn{2}{c}{2015} & \multicolumn{2}{c}{2016} \\
      & F &	M 	& F & M  &	F 	& M \\
     \hline \hline
     {\bf Participation in SDSS-III}:\\
     Leader and in SDSS-III         & 18	    & 43	& 21	    & 56	& 21	& 46 \\
     Leader and not in SDSS-III     & $<5$	    & 24	&   $<5$	& 29	&  7	& 15 \\
     Not Leader and in SDSS-III      & 15	    & 56	&  30       & 76	& 23	& 40 \\
     Not Leader and not in SDSS-III   & 27	    & 60	&  38	    & 86	& 36	& 47 \\
    \hline 
     {\bf Marital Status}:\\
     Leader and Single              & \nodata   & \nodata  & 14        & 67    & 20    & 44 \\
     Leader and Not Single          & \nodata   & \nodata  & 17        & 73    & 22    & 50 \\
     Not Leader and Single          & \nodata   & \nodata  & 25        & 43    & 19    & 20 \\
     Not Leader and Not Single      & \nodata   & \nodata  & 25        & 43    & 19    & 20 \\
    \hline
     {\bf Children}:\\
    Leader and No Children          & \nodata   & \nodata  & 14        & 39    & 16    & 23 \\
    Leader and Children             & \nodata   & \nodata  & 10        & 46    & 12    & 38 \\
    Not Leader and No Children      & \nodata   & \nodata  & 48        & 95    & 39    & 50 \\
    Not Leader and Children         & \nodata   & \nodata  & 19        & 67    & 21    & 38 \\
    \hline
    \end{tabular}
\end{table}



\end{document}